\documentclass[epj,nopacs]{svjour}
\usepackage{amsmath}
\usepackage{amssymb}
\usepackage[figuresright]{rotating}
\usepackage{color}
\usepackage{xspace}
\usepackage[gen]{eurosym}
\usepackage[mathscr]{eucal} 
\usepackage{latexsym}
\usepackage[left]{lineno}
\usepackage{graphics}
\newcommand{\nue}{\ensuremath{\nu_{e}}\xspace}

\newcommand{\numu}{\ensuremath{\nu_{\mu}}\xspace}

\newcommand{\nubarmu}{\ensuremath{\overline{\nu}_{\mu}}\xspace}

\newcommand{\boss}[2]{\ensuremath{\rlap{\kern-2.5pt\ensuremath{\overset{\scriptscriptstyle(-)}{\phantom{#1}}}}{\ensuremath{{#1}_{#2}}}}}

\newcommand{\bARII}              {INFN, Sezione di Bari, 70126 Bari, Italy}
\newcommand{\bARIU}            {Dipartimento di Fisica dell'Universit\`a  di Bari, 70126 Bari, Italy}
\newcommand{\bOLOGNAI}     {INFN, Sezione di Bologna, 40127 Bologna, Italy}
\newcommand{\bOLOGNAU}   {Dipartimento di Fisica e Astronomia dell'Universit\`a  di Bologna, 40127 Bologna, Italy}
\newcommand{\cERN}                  {European Organization for Nuclear Research (CERN), Geneva, Switzerland}
\newcommand{\fRASCATI}          {Laboratori Nazionali di Frascati dell'INFN, 00044 Frascati (Roma), Italy}
\newcommand{\lECCEI}             {INFN, Sezione di Lecce, 73100 Lecce, Italy}
\newcommand{\lECCEU}           {Dipartimento di Matematica e Fisica dell'Universit\`a  del Salento, 73100 Lecce, Italy}
\newcommand{\lECCEING}       {Dipartimento di Ingegneria dell'Innovazione, dell'Universit\`a  del Salento, 73100 Lecce, Italy}
\newcommand{\lEBEDEV}          {Lebedev Physical Institute of Russian Academy of Sciences, Leninskiy pr., 53, 119991 Moscow, Russia}
\newcommand{\lMEPhI}          {National Research Nuclear University MEPhI, 115409 Moscow, Russia}
\newcommand{\mSU}                 {Lomonosov Moscow State University (MSU SINP), 1(2) Leninskie gory, GSP-1, 119991 Moscow, Russia}
\newcommand{\pADOVAI}          {INFN, Sezione di Padova, 35131 Padova, Italy}
\newcommand{\pADOVAU}        {Dipartimento di Fisica e Astronomia dell'Universit\`a  di Padova, 35131 Padova, Italy}
\newcommand{\rOMAI}            {INFN, Sezione di Roma, 00185 Rome, Italy}
\newcommand{\zAGREB}             {Rudjer Boskovic Institute, Bijenicka 54, 10002 Zagreb, Croatia}
\newcommand{\bcINFN}     {INFN-CNAF, 40127 Bologna, Italy}
\begin{document}
%
\title{Search for Sterile Neutrinos in Muon Neutrino Disappearance Mode at FNAL}
\author{A.~Anokhina\inst{1},
A.~Bagulya\inst{2},
M.~Benettoni\inst{3},
P.~Bernardini\inst{4,5},
R.~Brugnera\inst{6,3},
M.~Calabrese\inst{5},
A.~Cecchetti\inst{7},
S.~Cecchini\inst{8},
M.~Chernyavskiy\inst{2},
F.~Dal~Corso\inst{3},
O.~Dalkarov\inst{2},
A.~Del~Prete\inst{5,9},
G.~De~Robertis\inst{10},
M.~De~Serio\inst{11,10},
D.~Di~Ferdinando\inst{8},
S.~Dusini\inst{3},
T.~Dzhatdoev\inst{1},
R.~A.~Fini\inst{10},
G.~Fiore\inst{5},
A.~Garfagnini\inst{6,3},
M.~Guerzoni\inst{8},
B.~Klicek\inst{12},
U.~Kose\inst{13},
K.~Jakovcic\inst{12},
G.~Laurenti\inst{8},
I.~Lippi\inst{3},
F.~Loddo\inst{10},
A.~Longhin\inst{3},
M.~Malenica\inst{12},
G.~Mancarella\inst{4,5},
G.~Mandrioli\inst{8},
A.~Margiotta\inst{14,8},
G.~Marsella\inst{4,5},
N.~Mauri\inst{8},
E.~Medinaceli\inst{6,3},
R.~Mingazheva\inst{2},
O.~Morgunova\inst{1},
M.~T.~Muciaccia\inst{11},
M.~Nessi\inst{13},
D.~Orecchini\inst{7},
A.~Paoloni\inst{7},
G.~Papadia\inst{5,9},
L.~Paparella\inst{11,10},
L.~Pasqualini\inst{14,8},
A.~Pastore\inst{10},
L.~Patrizii\inst{8},
N.~Polukhina\inst{2,15},
M.~Pozzato\inst{8},
M.~Roda\inst{6,3,}\thanks{\emph{Now at University of Liverpool, Department of Physics, Oliver Lodge
Laboratory, Liverpool L69 7ZE, UK}},
T.~Roganova\inst{1},
G.~Rosa\inst{16},
Z.~Sahnoun\inst{8,}\thanks{\emph{Also at Centre de Recherche en Astronomie Astrophysique et G\'eophysique, Alger, Algeria}},
T.~Shchedrina\inst{2},
S.~Simone\inst{11,10},
C.~Sirignano\inst{6,3},
G.~Sirri\inst{8},
M.~Spurio\inst{14,8},
L.~Stanco\inst{3},
N.~Starkov\inst{2},
M.~Stipcevic\inst{12},
A.~Surdo\inst{5},
M.~Tenti\inst{17},
V.~Togo\inst{8} \and
M.~Vladymyrov\inst{2}
}                     
%
%
\institute{\mSU \and 
\lEBEDEV \and 
\pADOVAI \and 
\lECCEU \and 
\lECCEI \and
\pADOVAU \and
\fRASCATI \and
\bOLOGNAI \and
\lECCEING \and
\bARII \and
\bARIU \and
\zAGREB \and
\cERN \and
\bOLOGNAU \and
\lMEPhI \and
\rOMAI \and
\bcINFN
}
%
%
\abstract{
\noindent The NESSiE Collaboration has been setup to undertake a conclusive experiment to clarify the  {\em muon--neutrino disappearance} measurements at short baselines 
in order to
put severe constraints to models with more than the three--standard neutrinos.
To this aim the current FNAL--Booster neutrino beam for a Short--Baseline experiment
was carefully evaluated by considering 
the use of magnetic spectrometers at two sites, near and far ones.
The detector locations were  studied, together with the achievable performances of two OPERA--like spectrometers.
The study was constrained by the availability of existing hardware and a time--schedule compatible with the undergoing project of multi--site Liquid--Argon detectors at FNAL.
\newline
The settled physics case and the kind of proposed experiment on the Booster neutrino beam would definitively clarify the existing tension between the $\numu$ disappearance and the $\nue$ appearance/disappearance at the eV mass scale. 
In the context of neutrino oscillations the measurement of $\numu$ disappearance is a robust and fast approach to either reject or discover new neutrino states at the eV mass scale. 
We discuss an experimental program able to  extend
by more than one order of magnitude (for neutrino disappearance) and by almost one order of magnitude (for antineutrino
disappearance) the present range of sensitivity for the mixing angle between standard and sterile neutrinos. These extensions
are larger than those achieved in any other proposal presented so far. 
\PACS{
      {PACS-key}{discribing text of that key}   \and
      {PACS-key}{discribing text of that key}
     } 
} 
\authorrunning{A.~Anokhina et al.}
\titlerunning{Search for sterile neutrinos in the \numu disappearance mode at FNAL}
\maketitle
\section{Introduction and Physics Overview}
\label{intro}
The unfolding of the physics of the neutrino is a long and pivotal history spanning the last 80 years. Over this period the interplay of 
theoretical hypotheses and experimental facts was one of the most fruitful for the progress in particle physics.
The achievements of the last decade and a half  brought out a coherent picture within the Standard Model (SM) or some minor extensions of it,
namely the mixing of three neutrino flavour--states with three  $\nu_1$, $\nu_2$ and $\nu_3$ mass eigenstates. 
Few years ago a non-vanishing $\theta_{13}$, the last still unknown mixing angle, was measured~\cite{theta13}.
Once the absolute masses of neutrinos, their Majorana/Dirac nature and the existence and magnitude of leptonic CP violation
be determined, the (standard) three--neutrino model will be {\em beautifully} settled. Still, other questions wou\-ld remain open: 
the reason for the characteristic nature of neutrinos, the relation between the
leptonic and hadronic sectors of the SM, the origin of Dark Matter and, overall, where and how to look for Beyond Standard Model (BSM) physics.
Neutrinos may be an excellent source of BSM physics and their history supports that possibility at length.

There are indeed several experimental hints for deviations from the ``coherent'' neutrino oscillation picture recalled above.
Many unexpected results, not corresponding to a discovery on a single basis, accumulated in the last decade and a half,
bringing attention to the hypothesis of the existence of {\em sterile neutrinos}~\cite{pontecorvo}. A White Paper~\cite{whitepaper}
provides a comprehensive review of these issues. 
In particular tensions in several phenomenological models grew up with experimental results on 
neutrino/antineutrino oscillations at Short--Baseline (SBL) and with the more recent, recomputed antineutrino--fluxes from 
nuclear reactors. 

\begin{figure}[htbp]
\includegraphics[width=9cm]{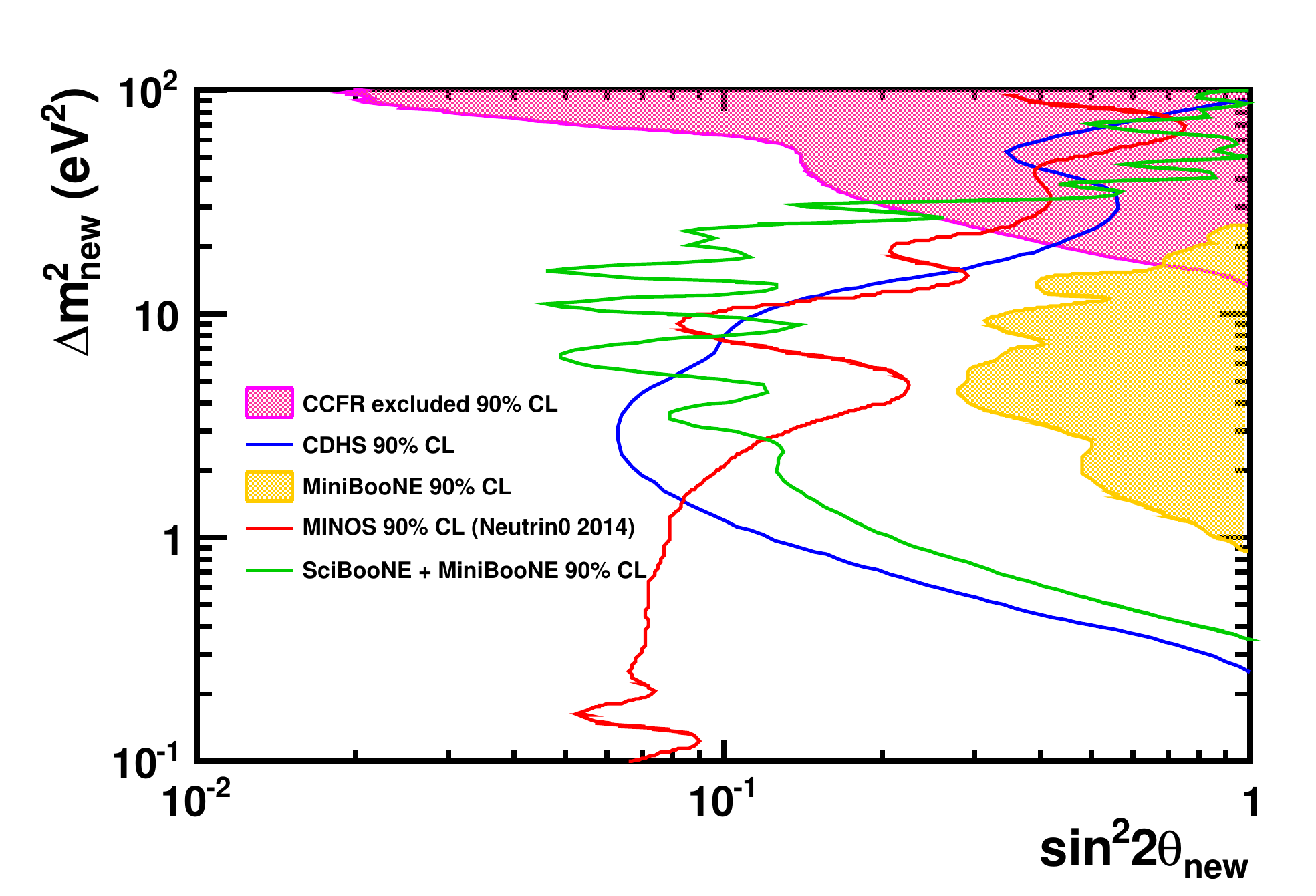}%
\caption{\label{fig:old-res}(color online) The current exclusion limits on the \numu disappearance searches at the eV$^2$ scale.
Blue (green) line: old (recent) exclusion limits on \numu from previous CDHS~\cite{CDHS} and recent MiniBooNE/SciBooNE~\cite{mini-sci-mu} measurements.
The two filled areas correspond to the exclusion limits on the \nubarmu from CCFR~\cite{ccfr} and MiniBooNE--alone~\cite{mini-mu} experiments.
The red curve corresponds to the very recent result from MINOS~\cite{recent-MINOS}. All the exclusion limits were compute at 90\% C.L..
}
\end{figure}

The main source of tension originates from the absence so far of any \numu 
disappearance signal~\cite{kopp-tension}.
Limited experimental data are available on searches for \numu disappearance at SBL: the rather old CDHS experiment~\cite{CDHS} and
the more recent results from MiniBooNE~\cite{mini-mu}, a joint MiniBooNE/SciBooNE analysis~\cite{mini-sci-mu} and the 
MINOS~\cite{recent-MINOS} and SuperKamiokande~\cite{recent-SK} exclusion limits reported at the NEUTRINO2014 conference.
The tension between \nue appearance and \numu disappearance was actually strengthened by the
MINOS and SuperKamiokande results, even if they only slightly extend the $\nu_\mu$ disappearance exclusion region set previously mainly by CDHS and
at higher mass scale by the CCFR experiment~\cite{ccfr}.
Fig.~\ref{fig:old-res} shows the excluded regions in the parameter space that describe SBL \numu disappearance induced by a sterile neutrino. 
The mixing angle is denoted as $\theta_{new}$ and the squared mass difference as
$\Delta m^2_{new}$. As evident from Fig.~\ref{fig:old-res}, the region $\sin^2(2\theta_{new})<0.1$ is still largely 
unconstrained. 
While this paper was being processed by the referees of the Journal new results were made available. In particular
a joint analysis by MINOS and DAYA-BAY~\cite{new-ster-1}, 
and the IceCube experiment~\cite{new-ster-2}.
Their results exclude part of the phase space $\sin^2(2\theta_{new})<0.1$ even if the critical region $\Delta m^2_{new}\sim1$ eV$^2$
is still only marginally touched,
while the \numu -- \nue
tension from global--fits stays around 0.04--0.07 for $\sin^2(2\theta_{new})$~\cite{kopp-tension}. For \nubarmu the situation is even worse as it will be 
further discussed in Section~\ref{sec:antinu}.

The outlined scenario promoted several proposals for new, exhaustive evaluations of the neutrino phenomenology
at SBL. Since the end of 2012 CERN started the setting up of a {\em Neutrino Platform}~\cite{edms}, with new
infrastructures at the North Area that, for the time being, does not include a neutrino beam. Meanwhile in the US, FNAL 
welcomed proposals for experiments exploiting the physics potentials of their two existing neutrino beams, the Booster and the NuMI beams, following the
 recommendations from USA HEP--P5 report~\cite{p5}.
Two  proposals~\cite{ICARUSFNAL,LAr1-ND} were submitted for SBL experiments at the Booster beam, to complement
the about to start MicroBooNE experiment~\cite{microboone}. They 
are all based on the Liquid--Argon (LAr) technology and aim to measure the \nue appearance at SBL, with less possibilities to 
investigate the \numu disappearance~\cite{SBN}. In this paper a complementary case study based on magnetic spectrometers at two different sites at FNAL--Booster beam is discussed, built up on the following considerations:
\begin{enumerate}
\item the measurement of \numu \footnote{From hereafter \numu refers to
either \numu or \nubarmu, unless otherwise stated.}
spectrum in both normalization and shape is mandatory for a correct interpretation of the \nue data, even in case of 
a null result for the latter;
\item a decoupled measurement of \nue and \numu interactions allows to reach in the analyses the percent--level systematics due to the 
different cross--sections;
\item {\em very massive detectors} are mandatory to collect a large number of events thus improving the disentangling
of systematic effects.
\end{enumerate}

This paper is organized in the following way. After the introduction a short overview of the NESSiE proposal is given. A detailed report of the  studies  performed
on the constraints of the FNAL--Booster neutrino beam is drawn in Section~\ref{sec:beam}. In Section~\ref{sec:spect1} a description of the detector system and the corresponding outcomes are provided. 
The statistical analyses and the attainable exclusion-limits on the \numu and \nubarmu disappearances are depicted
in Sections~\ref{sec:analysis} and~\ref{sec:antinu}, respectively. Finally conclusions are drawn.


\section{Proposal for the FNAL--Booster beam}\label{expo}

Assuming the use of the FNAL--Booster neutrino beam a detailed study of the physics case  was performed along the lines 
followed when considering neutrino beams at CERN--PS and CERN--SPS~\cite{nessie,larnessie} and the approach of the analysis reported in~\cite{stancoetal}. A substantial difference between FNAL and CERN beams is the 
decrease of the average neutrino energy by more than a factor 2,
thus making the study very challenging for an high Z--density detector. 
Several detector configurations were studied, investigating experimental aspects not fully addressed by the LAr detection, such as 
the measurements of the lepton charge on event--by--event basis and the lepton energy over a wide range.
Indeed, muons from Charged Current (CC) neutrino interactions play an important role in disentangling 
different phenomenological scenarios provided their charge state is determined. Also,
the study of muon ap\-pea\-ran\-ce/dis\-ap\-pea\-ran\-ce can benefit from the large statistics of CC events from the primary muon
neutrino beam. 

In the FNAL--Booster neutrino beam the antineutrino contribution is rather small and it corresponds to a systematic effect to be taken into account.  For the antineutrino beam the situation is rather different since a large flux
 of neutrinos is also present.
From an experimental perspective the possibility of an event--by--event
detection of the primary muon charge is an added value since it allows
to disentangle the presence/absence of new effects which might
genuinely affect antineutrinos and neutrinos differently (CP
violation is possible in models with more than one additional
sterile neutrino). This possibility is particularly intriguing while running
in negative horn polarity due to the sizable contamination from the
cross-section enhanced interactions of parasitic neutrinos.

The extended NESSiE proposal is available in~\cite{nessie-fnal}.  It consists in the
design, construction and installation of two spectrometers at two sites,
{\em Near} (at 110 m, on--axis) and {\em Far} (at 710 m, off-axis, on surface), in line with
the FNAL--Booster beam and compatible with the proposed LAr detectors. 
Profiting of the large mass of the two spectrometer--systems, their performances as stand--alone apparatus are exploited for the \numu 
disappearance study. Besides, complementary measurements with the foreseen LAr--systems can be undertaken to increase their control of
systematic errors.

Practical constraints were assumed in order to draft a proposal on a
conservative, manageable basis, with sustainable timescale and cost--wise.
Well known technologies were considered as well as re--using of large parts of existing detectors. 

The momentum and charge state measurements of muons in a wide range, from few hundreds MeV/c to several 
GeV/c, over a $> \ 50\ {\rm m}^2$ surface, are an extremely challenging task.
In the following, the key features of the proposed experimental layout are presented. 
By keeping the systematic error at the level of $1-2$\% for the detection of the \numu interactions, it will be possible to:
\begin{itemize}
	\item
measure the \numu disappearance in a large muon--mo\-men\-tum, $p_{\mu}$, range (conservatively a $p_{\mu}\ge 500$~MeV/c cut is chosen) in order to
reject existing anomalies over the whole expected parameter space of sterile neutrino oscillations 
at SBL;
\item collect a very large statistical sample so as
to test the hypothesis of muon (anti)neutrino disappearance for values of the 
mixing parameter down to still un--explored regions ($\sin^2(2\theta_{new})\lesssim 0.01$);
	\item
measure the neutrino flux at the near detector, in the relevant muon momentum range, in order to keep the systematic errors at the lowest possible values;
\item measure the sign of the muon charge to
separate \numu from \nubarmu for the control of the systematic error.
\end{itemize}


\section{Beam evaluation and constraints}\label{sec:beam}

For a proposal that aims to make measurements with the FNAL--Booster muon--neutrino beam the convolution of the beam features 
and of the muon detection constitutes the major constraint. An extended study was therefore performed.

\subsection{The Booster Neutrino Beam (BNB)}

The neutrino beam~\cite{G4BNBflux} is produced from protons with a
kinetic energy of 8 GeV extracted from the Booster and directed to a
Beryllium cylindrical target 71~cm long and with a 1~cm diameter.
The target is surrounded by a magnetic focusing horn pulsed
with a 170~kA current at a rate of 5~Hz. Secondary mesons are
projected into a 50~m long decay--pipe where they are allowed to decay
in flight before being stopped by an absorber and the ground
material.  An additional absorber could be placed in the decay pipe at
about 25~m from the target. This configuration, not currently
in use, would modify the beam properties providing a
more point--like source at the near site and thus extra
experimental constraints on the systematic errors.  

Neutrinos travel
about horizontally at a depth of 7~m underground.
Proton batches from the Booster contain $\sim 4.5\times 10^{12}$
protons, have a duration of 1.6 $\mu$s and are subdivided into 84
bunches. Bunches are $\sim 4$~ns wide and are separated by 19~ns. The
rate of batch extraction is limited by the horn pulsing at 5 Hz. This
timing structure provides a very powerful constraint to the background
from cosmic rays.



\subsection{The Far--to--Near Ratio (FNR)}\label{sec:fnr-ratio}


\begin{figure}[htbp]
\centering
\includegraphics[scale=0.445]{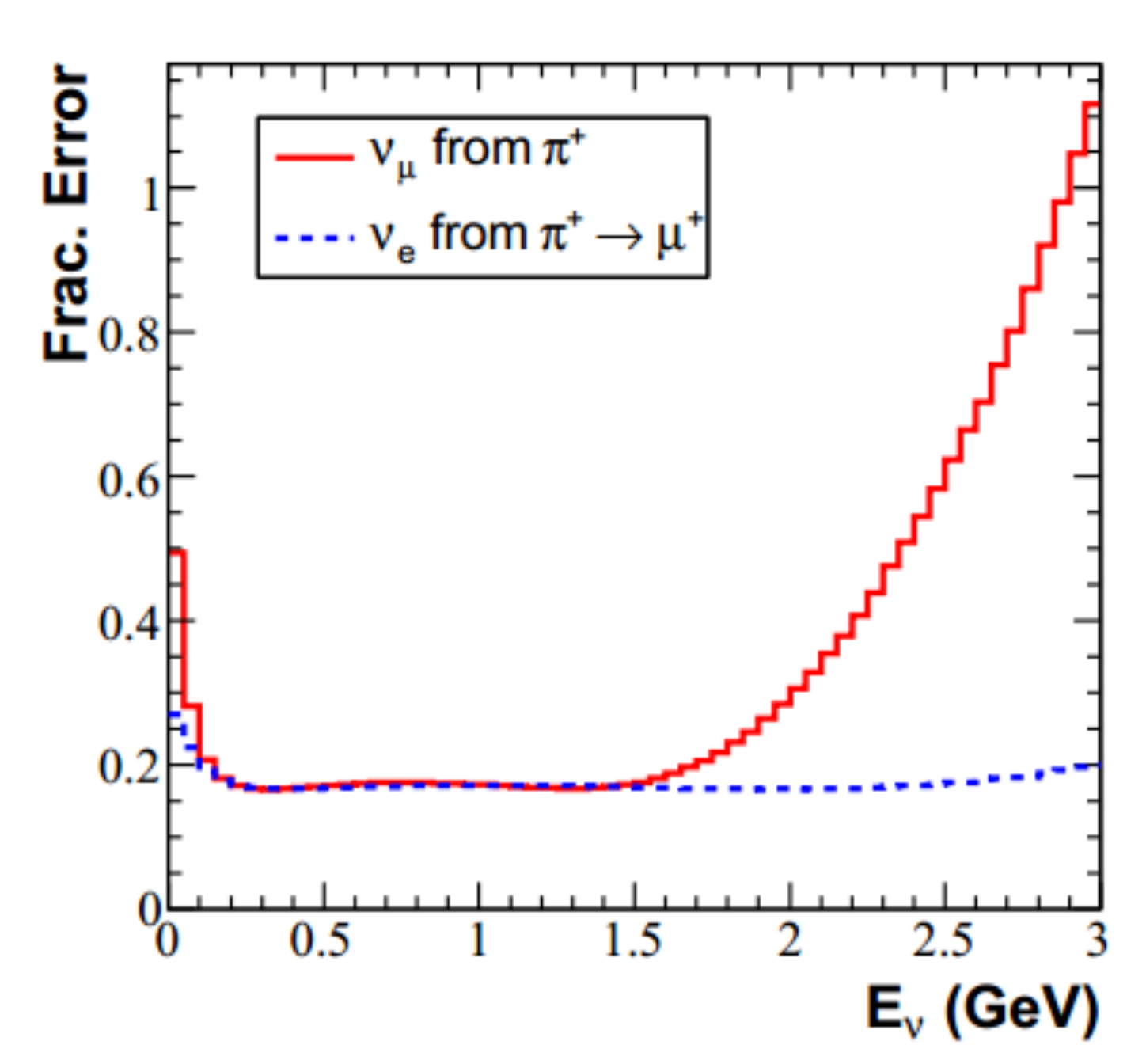}\\
\includegraphics[scale=0.25]{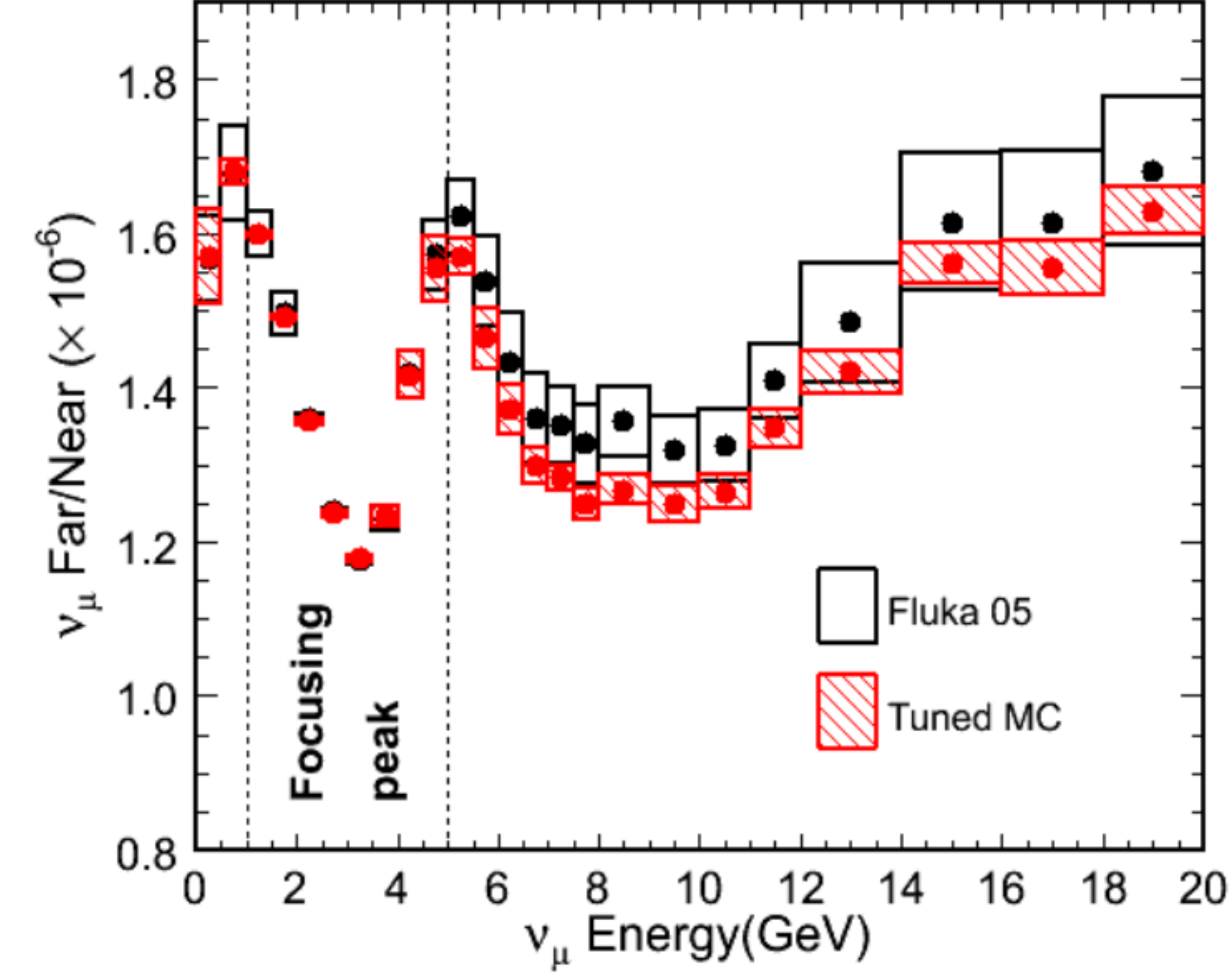}
\caption{\label{fig:hpmb}(color online) Top: uncertainties on the absolute flux of $\nu_\mu$ at MiniBooNE (from~\cite{G4BNBflux}).
Bottom: uncertainties on the far--to--near ratio at NuMI according to different simulations (from~\cite{numikopp}).}
\end{figure}


The uncertainty on the absolute $\nu_\mu$ flux at MiniBooNE, shown
in Fig.~\ref{fig:hpmb}, top (from~\cite{G4BNBflux}), stays below
20\% for energies below 1.5~GeV, increasing drastically at larger energies and also below 200 MeV.
The uncertainty is dominated by the knowledge of proton interactions
in the Be target, which affects the angular and momentum
spectra of neutrino parents. The result of
Fig.~\ref{fig:hpmb} is based on experimental data obtained by the HARP
and E910 collaborations~\cite{G4BNBflux}.

The large uncertainty on the absolute neutrino flux makes the use of two or more identical
detectors at different baselines mandatory when searching for small
disappearance phenomena. The ratio of the event rates at the far and
near detectors (FNR) as function of neutrino energy is a convenient
variable since at first order it benefits from cancellation of
systematics due to the common effects of proton--target, neutrino
cross--sections and reconstruction efficiencies.
Because of these cancellations the uncertainty on the FNR or,
equivalently, on the spectrum at the far site extrapolated from the spectrum at the near site
is at the level of few percent. As an example the FNR for the NuMI
beam is shown in bins of neutrino energy in Fig.~\ref{fig:hpmb}, bottom (from~\cite{numikopp}); the
uncertainty ranges in the interval 0.5--5.0\%.

It is worth to note that, even in the absence of oscillations, the energy
spectra in any two detectors are different, thus leading to a non--flat
FNR. This is especially true if the distance of the near detector is
comparable with the length of the decay pipe. It is therefore necessary
to master the knowledge of the FNR for physics searches.

\begin{figure}[htbp]
\centering
\includegraphics[scale=0.35]{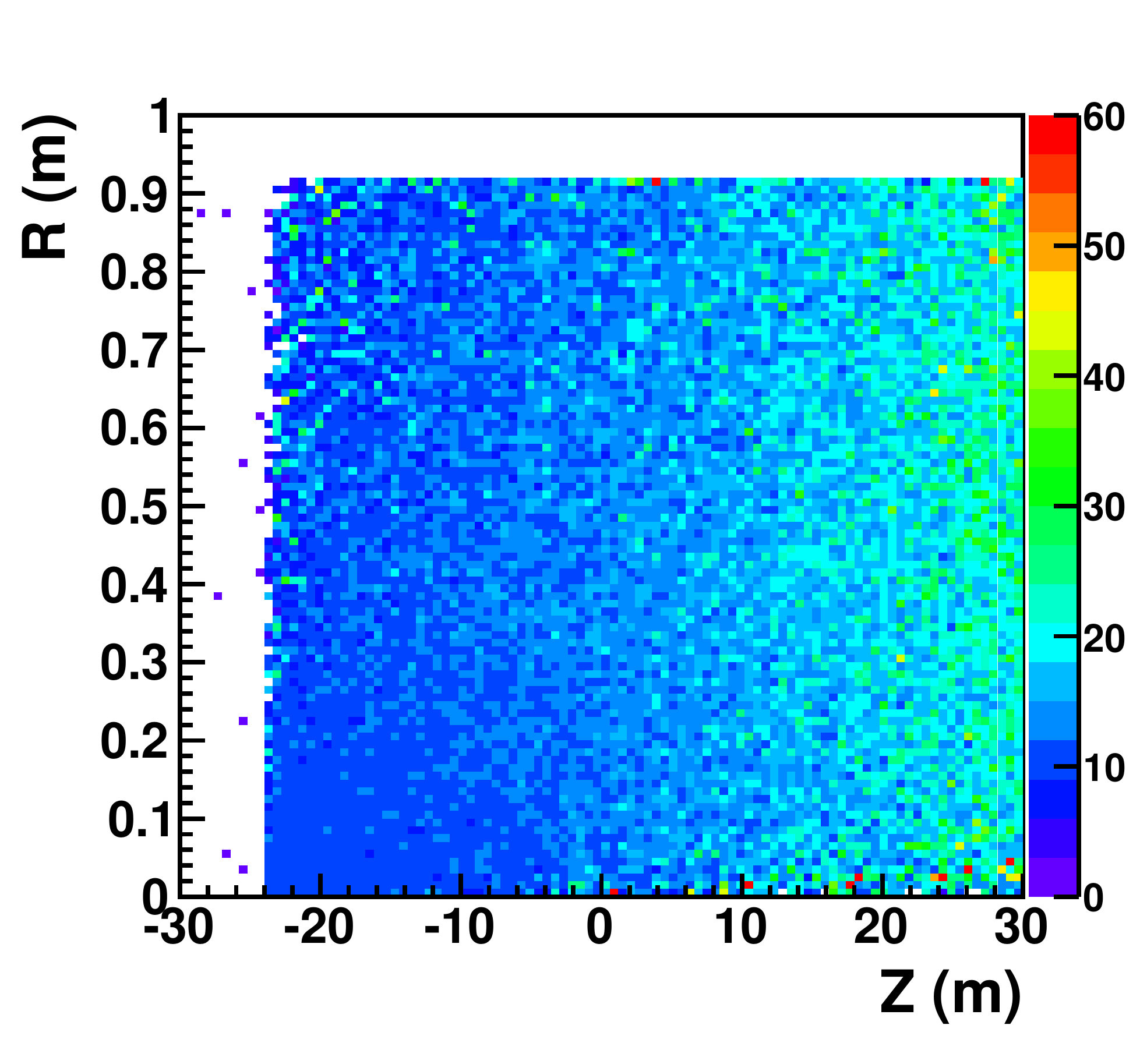}
\caption{\label{fig:teff}(color online) Ratio between the $Z$--$R$ distributions of neutrino
  production points for neutrinos crossing  a near--detector transverse surface of $4\times 4$ m$^2$ at 110~ m  over
  neutrinos crossing a far--detector transverse surface of $8\times 8$ m$^2$ at 710~ m.  The $Z$ origin is fixed at the middle of the decay pipe,
  $R$ being the radial distance in the pipe.
  The near detector has a
  larger acceptance for neutrinos produced in the most downstream part
  of the decay pipe, as expected.}
\end{figure}

Assuming a transverse area for the detectors at near and far sites of the same order,
the solid angle subtended by the near detector is larger than that subtended by the far one.
Therefore, neutrinos, and mostly those from mesons decaying at the end of the pipe,
have a higher probability of being detected in the near than in the far detector.  
In the far detector, on the contrary, only neutrinos produced in a narrow forward cone
are visible.
This effect 
is illustrated in Fig.~\ref{fig:teff} showing the ratio of the integrated neutrino flux at the two locations
distributed over the neutrino production points (radius $R$ vs
longitudinal coordinate $Z$), for a sample crossing a near ($4\times 4$ m$^2$ transverse area) and a far
($8\times 8$ m$^2$) detector with front--surface placed at 110 and 710 m from the target, respectively.
Neutrinos produced at large $Z$ can be detected even if they are
produced at relatively large angles, enhancing the contribution of lower energy neutrinos.
On the other hand neutrinos from late decays come from the fast pion component that is more
forward--boosted. The former effect is the leading one so the net
effect is a softer spectrum at the near site.


\begin{figure}[htbp]
\centering
\includegraphics[scale=0.445,type=pdf,ext=.pdf,read=.pdf]{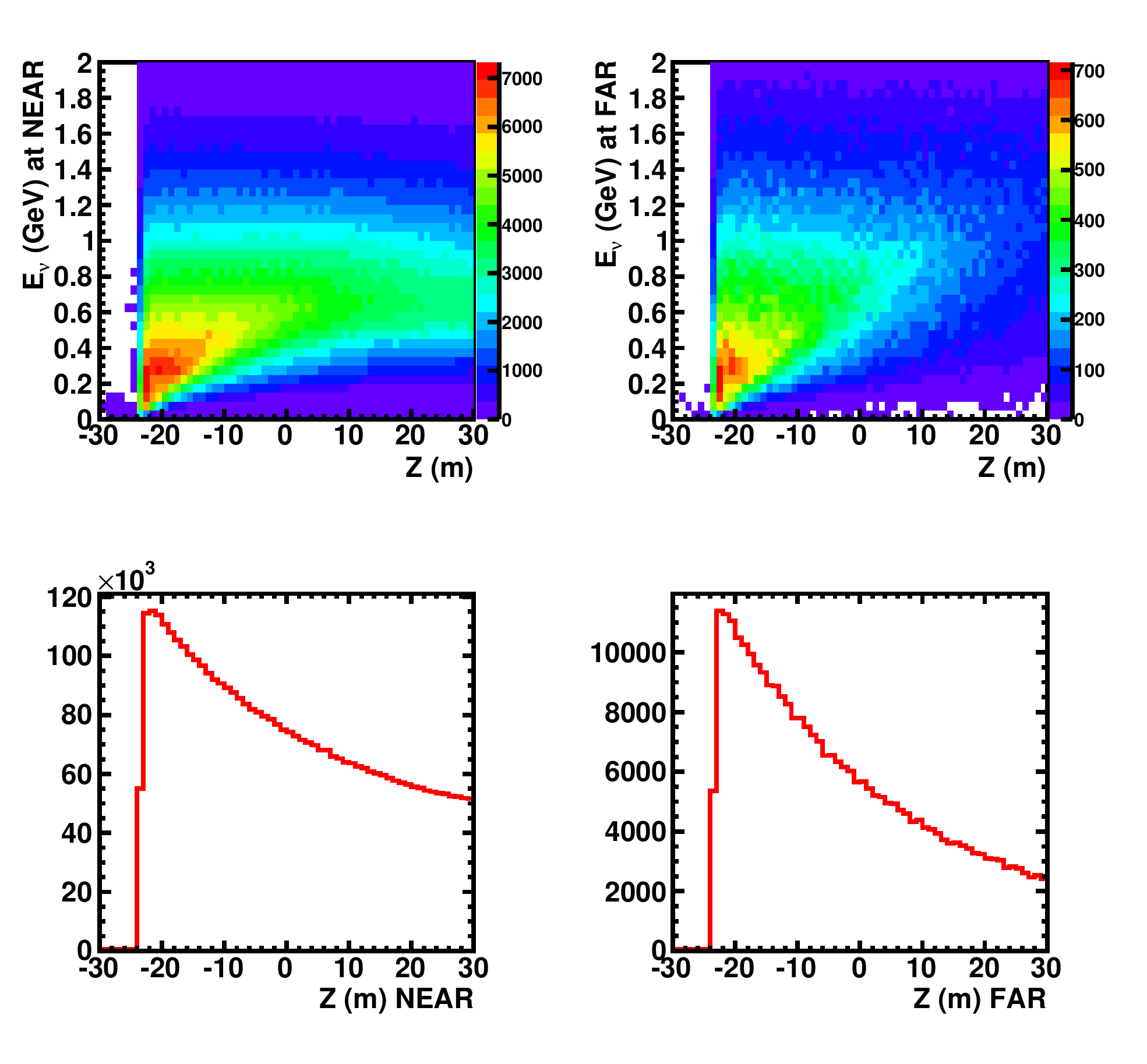}
\caption{\label{fig:teff1}(color online) Distribution of $E_\nu$ vs $Z$ for neutrinos seen in the near (top left)
and far (top right) detector. Distribution of the $Z$ production points for neutrinos seen in the near 
(bottom left) and far (bottom right) detectors. The assumed detector active surface is a square of $4 \times 4$~m$^2$ 
and $8 \times 8$~m$^2$ for
the near and far detector, respectively.}
\end{figure}

In Fig.~\ref{fig:teff1} (top plots) the distributions of the neutrino energy, $E_\nu$, vs
$Z$ for neutrinos crossing the near (top left) and the far site (top
right) are shown. The assumed detector active surface is a square of $4 \times 4$~m$^2$ 
and $8 \times 8$~m$^2$ for
the near and far detector, respectively.
As anticipated, the energy spectrum at the near site is
softer, the additional contribution at low energy being particularly
important for neutrinos from late meson--decays.  The
distribution of $Z$ is also shown in Fig.~\ref{fig:teff1} for
neutrinos crossing the detectors at the near (bottom left) and far (bottom right) sites.


From these considerations it is apparent that the
prediction of the FNR is a delicate task requiring the full simulation
of the neutrino beam--line and the detector acceptance. Moreover, the systematic uncertainties 
on the FNR parameter play a major role requiring deep investigation.


The various contributions to the systematic uncertainties on the neutrino flux were studied in detail by the MiniBooNE collaboration 
in~\cite{G4BNBflux} (Tab.~\ref{tab:uncertainties}). At first order they factorize out using a double site. However, 
since their magnitude can limit the FNR accuracy, we studied in detail the largest contribution, which comes from the knowledge
of the hadro--production double differential (momentum $p$, polar angle
$\theta$) cross--sections in 8 GeV $p$--Be interactions.



\begin{table}
\caption{\label{tab:uncertainties}Systematic uncertainties on the flux prediction of the $\nu_\mu$ Booster beam.}
\centering
\begin{tabular}{lr}
\hline
source &  error (\%) \\
\hline
$p$--Be $\pi^+$ production &14.7\\
2$^{ry}$ nucleons interaction &2.8\\
$p$--delivery & 2.0\\
2$^{ry}$ pions interaction &1.2\\
magnetic field & 2.2\\
beam--line geometry & 1.0\\
\hline
\end{tabular}
\end{table}



Other contributions are less relevant and do not practically affect the FNR estimator. As an example the systematic contributions
due to the multi-nucleon and the final state interactions have been investigated. 
Their modeling can be important when measuring 
cross-sections or for the extraction of oscillation parameters with measurements from a single detector. 
However, the local interaction is the same when two sites, near (N) and far (F), are used. Any estimator 
of the F/N ratio in terms of some measurable quantity correlated to the neutrino energy is not affected  at first order by shape
distortion. Therefore, in case of near and far detections 
the effect of the interaction models becomes sub-leading. 
What matters is the convolution of the neutrino interaction model with fluxes,
detector acceptance as well as detector composition, which may be different at the two sites. 
The amount of this sub-leading contribution depends also on the characteristic of the detectors: 
Water Cherenkov, Liquid Argon, Scintillator, Iron etc.

\begin{figure}[htbp]
\centering
\includegraphics[scale=0.35,type=pdf,ext=.pdf,read=.pdf]{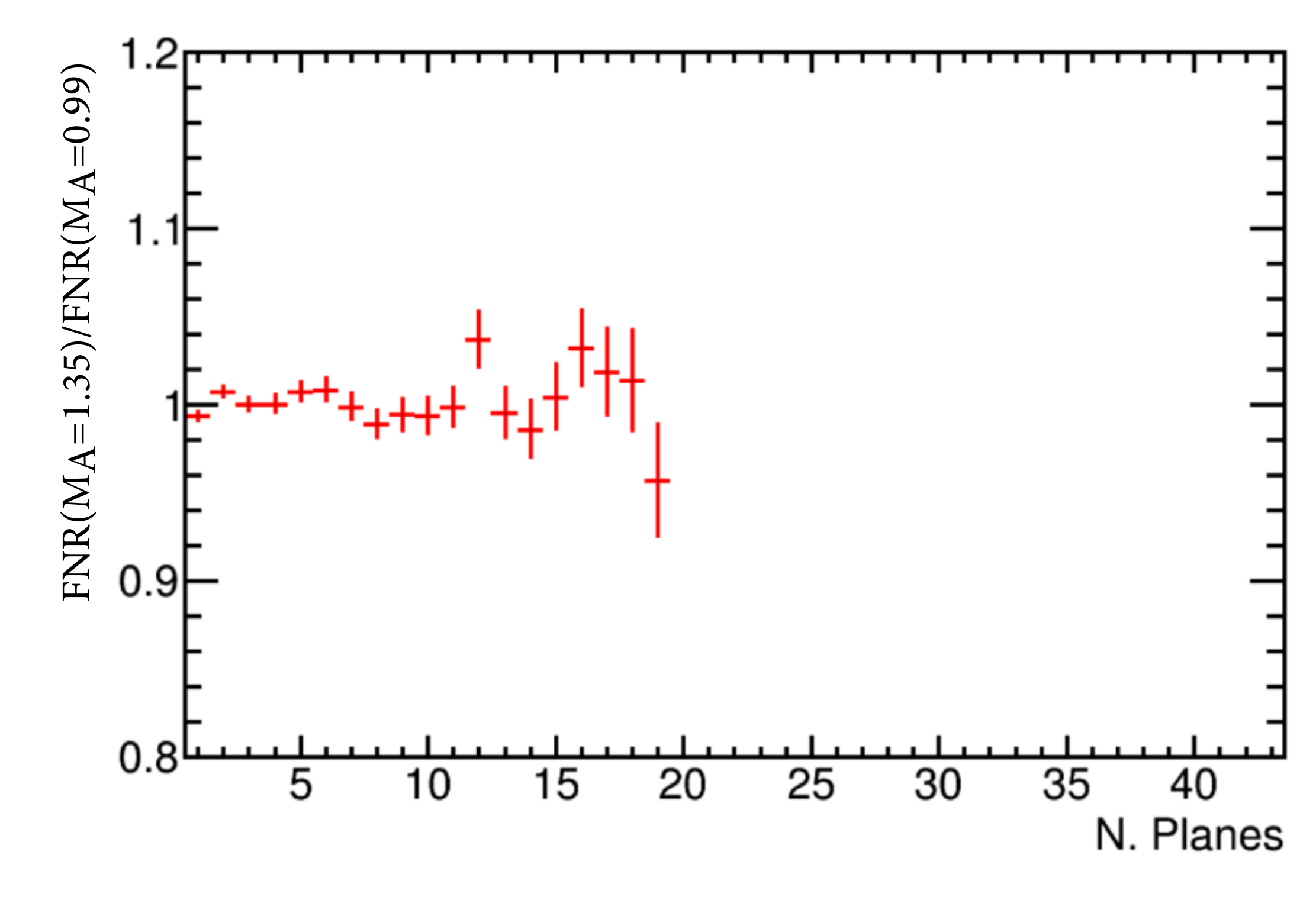}
\caption{\label{fig:unc-int}(color online) The FNR systematic distortion due to different modeling of the neutrino
interaction. An extreme value has been chosen for the axial-mass, $M_A=1.35$~GeV, compared to the standard one,
$M_A=0.99$~GeV. The distortion is shown as function of the crossed iron planes of the detector (see Section~\ref{sec:spect1});
it corresponds to the systematic error due to the convolution of the neutrino interaction, the fluxes and the detectors' acceptance.}
\end{figure}

NESSiE is the only proposal that could plainly profit of its identical configuration at near and far
sites (up to the iron composition of the corresponding slabs in the near and far detectors), its
capability to contain the events, and the control of the F/N ratio in various muon momentum
ranges with large statistics.
The contribution of the interaction models to the systematic error of  
the F/N ratio was checked and found very small (below 1\%).
An example of the performed several checks  is reported in Fig.~\ref{fig:unc-int} and Tab.~\ref{tab:unc-int}, for two extreme cases of the axial-mass, $M_A=0.99$ and 
$M_A=1.35$~GeV, analyzed with a full chain simulation (GENIE~\cite{genie}/FLUKA plus GEANT4 applied to 
configuration 4, see next section).

\begin{table}[htbp]
\caption{\label{tab:unc-int}Systematic uncertainties on the FNR estimator as function of the crossed iron planes due to 
the convolution of the neutrino interaction, the fluxes and the detectors' acceptance, as shown
in Fig.~\ref{fig:unc-int}, up to a muon momentum of about 1 GeV. Uncertainties average at 0\% with a spread  $<$ 1\%
due to the discreetness of the variable used.}
\centering
\begin{tabular}{cc cc cc}
\hline
nb. pl. &  (\%) & nb. pl. &  (\%) & nb. pl. &  (\%) \\
\hline
0 & -0.63 & 4 & 0.7 & 8 & -0.5\\
1 & 0.7 & 5 & 0.9 & 9 & -0.6\\
2 & 0 & 6 & -0.1 & 10 & -0.1\\
3 & 0.1 & 7 & -1.1 & 11 & 3.7\\
\hline
\end{tabular}
\end{table}

\subsection{Monte Carlo beam simulation}

In order to evaluate how the hadro--production uncertainty affects the
knowledge of the FNR in our experiment
a new beam--line simulation was developed.
The angular and momentum distribution of pions exiting the Be target
were simulated using either FLUKA (2011.2b)~\cite{fluka} or GEANT4 (v4.9.4
p02, QGSP 3.4 physics list). Furthermore the Sanford--Wang
parametrization for $\pi^+$ determined from a fit of the HARP
and E910 data--set in~\cite{G4BNBflux}, was used:
\begin{equation}
\frac{d^2\sigma}{dpd\Omega} = c_1 p^{c_2}\left(1-\frac{p}{p_b-1}\right)e^{-\frac{p^{c_3}}{p_{b}^{c_4}}-c_5\theta(p-c_6p_b\cos^{c_7}\theta)}
\label{eq:SW}
\end{equation}
with $p_b$ being the proton--beam momentum and $c_i$ ($i=1\ldots$ 7) free parameters.
The additional subdominant contributions arising
from $\pi^-$ and $K$ decays have been neglected when considering positive
polarity beam configurations.

For the propagation and decays of secondary mesons a
simulation using GEANT4 libraries was developed. A simplified version of the beam--line
geometry was adopted.  Despite the
approximations a fair agreement with the official simulation
of the MiniBooNE Collaboration~\cite{G4BNBflux} was obtained. This tool is
sufficient for the purpose of the site optimization that is
described in the following. In order to fully take  into account
finite--distance effects, fluxes and spectra were derived after
extrapolating neutrinos down to the detector volumes without using
weighing techniques. A total number of $7 \times 10^8$ protons on
target (p.o.t.), $2.1\times 10^8$ p.o.t. and $1 \times 10^9$ pions were
simulated with FLUKA, GEANT4 and Sanford--Wang parametrization,
respectively.



\begin{figure}
\centering
\includegraphics[scale=0.5,type=pdf,ext=.pdf,read=.pdf]{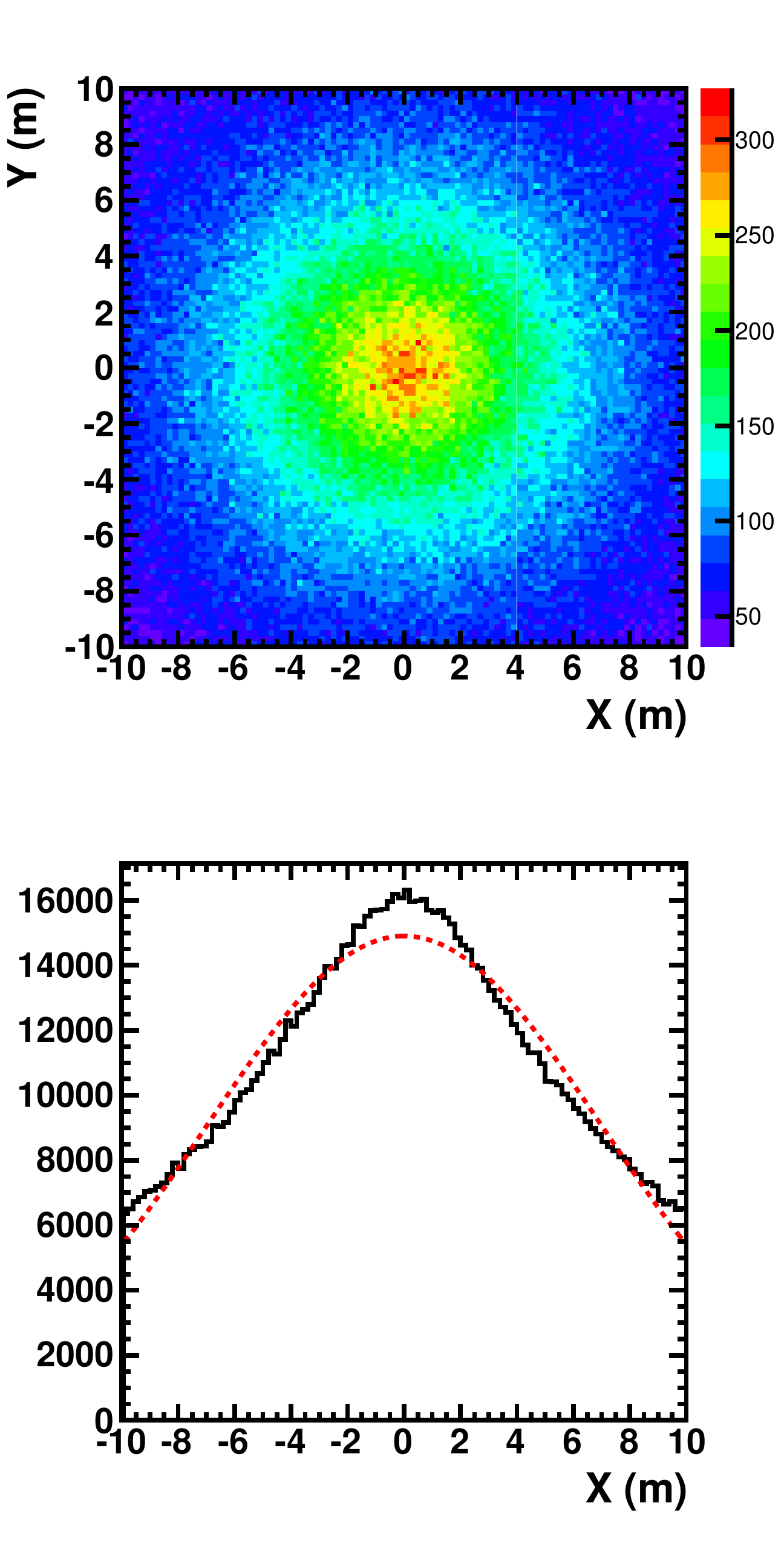}
\caption{\label{fig:fig_1}(color online) The neutrino beam profile at 110 m from the target. In the bottom plot the projection on the horizontal axis, $X$,  fitted 
 to a Gaussian curve for comparison.}
\end{figure}

In Fig.~\ref{fig:fig_1}  the transverse distributions of neutrinos at a distance of 110~m from the
target is shown. The root--mean--square (r.m.s.) of the distribution is about 5~m. The projected coordinate is shown in the bottom
plot of Fig.~\ref{fig:fig_1} with a Gaussian fit superimposed for comparison. The plot indicates that a
 near detector placed on ground--surface ($Y=7$ m) would severely limit the statistics (furthermore  the
angular acceptance of the far and near detectors would be too different).


\subsection{Choice of experimental sites}

Once the geometry and the mass of the detectors have been fixed additional
issues affect the choice of the location of the experimental sites, near and far ones. 
The ultimate figure of merit is the power of exclusion (or discovery) 
for effects induced by sterile neutrinos in a range of parameters as 
wide as possible in a given running time\footnote{A similar optimization process aimed to find the best
location in front of the Booster beam was extensively performed by the SciBooNE collaboration~\cite{sci-opt}.
In that case the aim was either to maximize the neutrino flux or to shape out the energy interval
for cross--section measurements.}. 

As soon as the detectors are further away from the target they ``see'' more similar spectra since the production region 
better approximates a point--like source. This helps in reducing the systematic uncertainty.
On the other hand the larger is the distance the smaller is the size of the collected event sample. Moreover the 
lever--arm for oscillation studies is reduced.
The reliability of the simulation of the neutrino spectra at the near and far sites remains an essential condition. 
This point is further addressed in Section~\ref{subsectsysconf}.

On a practical basis the increasing of the depth of the detector sites impacts considerably on the civil engineering
costs. 
Furthermore existing or proposed experimental facilities
(SciBooNE/LAr1--ND, T150--Icarus, MiniBooNE, MicroBooNE, LAr1, Icarus) already
partially occupy possible sites along the beam line~\cite{SBN}.


\subsubsection{Dependence of $\nu_\mu^{CC}$ rates and energy spectra on the detector position}

\begin{figure}
\centering
\includegraphics[scale=0.4,type=pdf,ext=.pdf,read=.pdf]{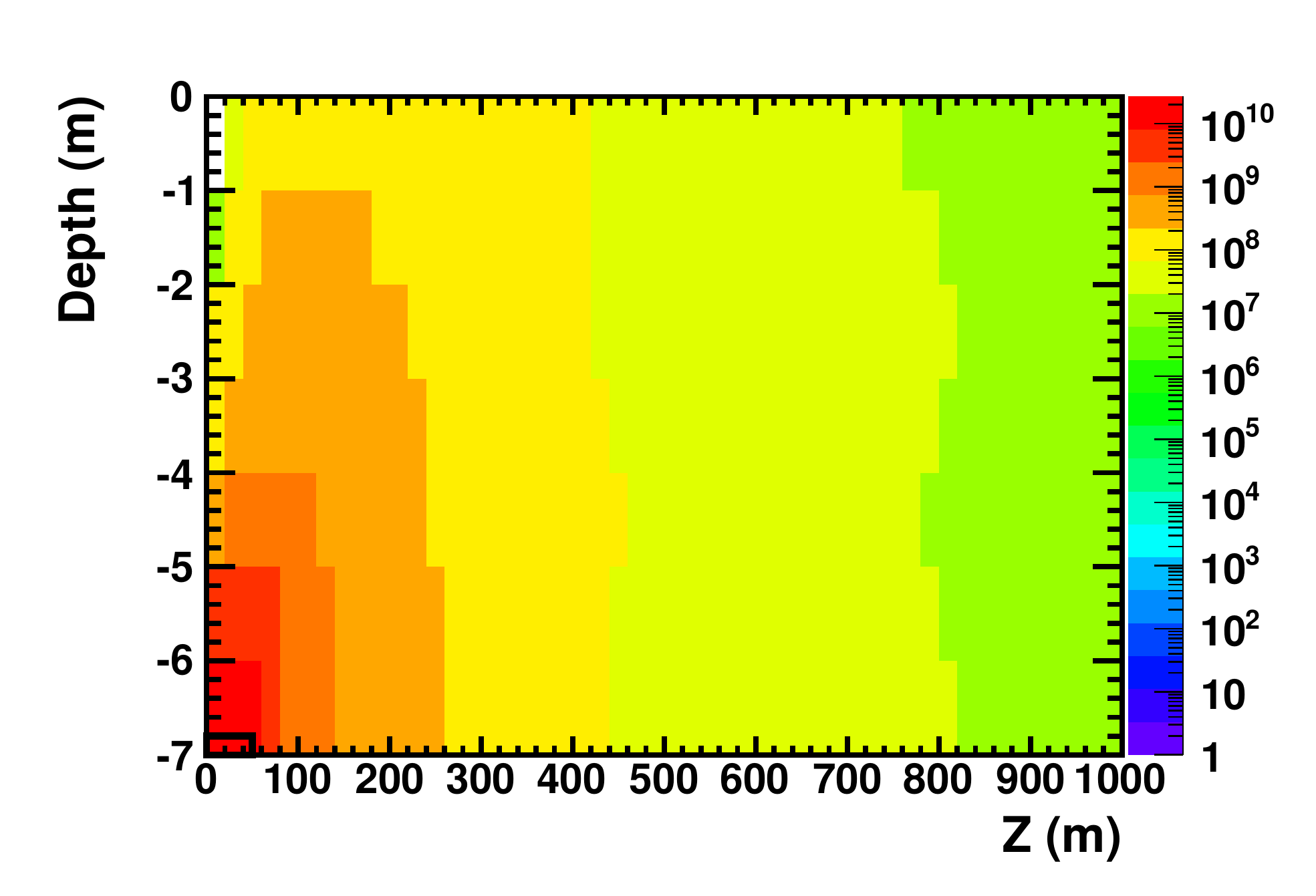}\\
\includegraphics[scale=0.4,type=pdf,ext=.pdf,read=.pdf]{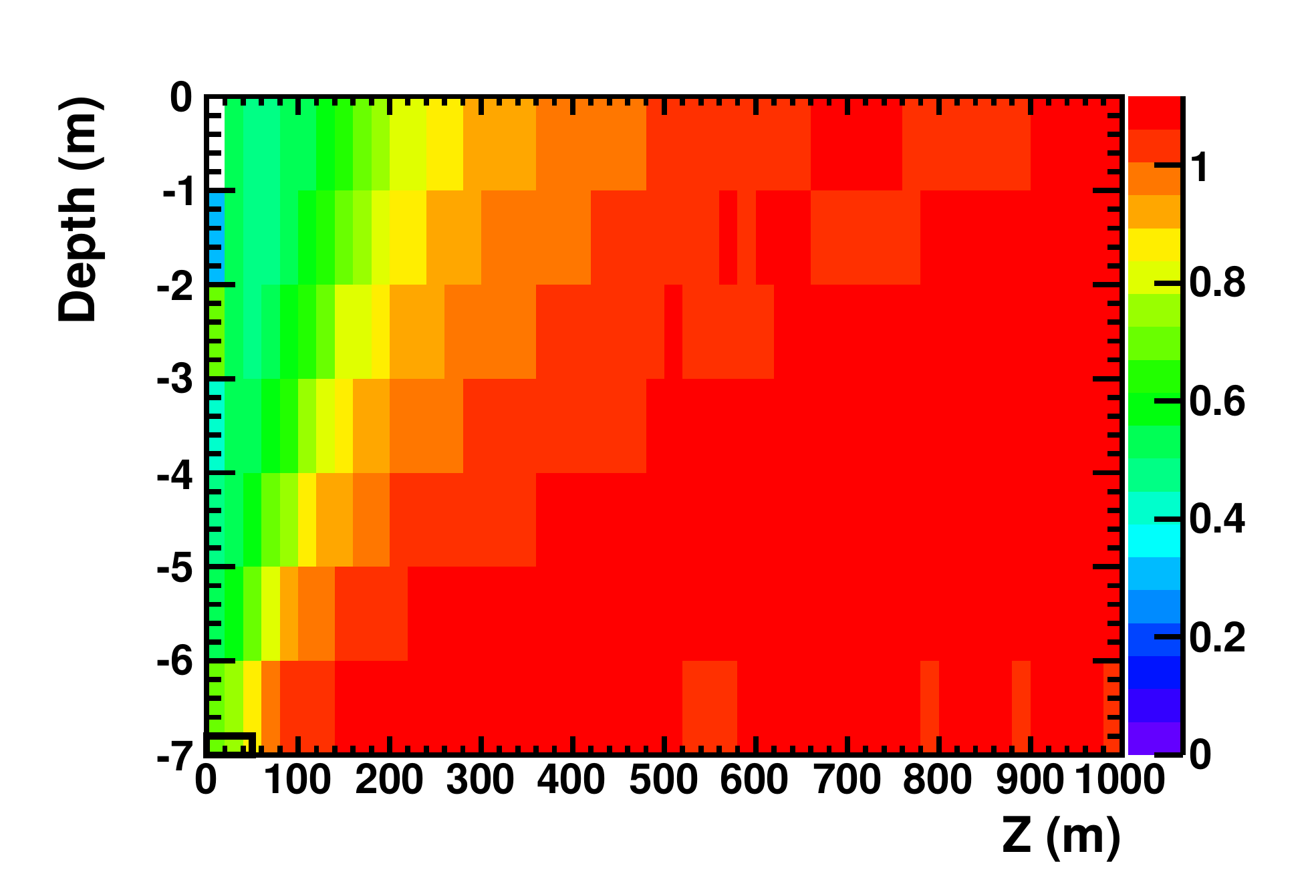}
\caption{\label{fig:figscan}(color online) Top: $\nu_\mu^{CC}$ rate (a.u.) vs distance
from the target,  $Z$, and the depth from the ground surface. Bottom: mean energy (in GeV) of $\nu_\mu^{CC}$ interactions vs $Z$ (m) and depth.
The origin of the reference system is the proton--target upstream position.
The hollow box close to the origin indicates the longitudinal extension of the decay pipe.}
\end{figure}

The $\nu_\mu^{CC}$ interaction rates and their mean energy depend on the
distance from the proton target, as shown in Fig.~\ref{fig:figscan}, top and bottom, respectively. 
The horizontal axis corresponds to
the distance ($Z$) from the target,
the vertical axis to the depth from the ground surface.
At a distance of about 700 m the rates and the mean energies
are barely affected when moving from on--axis to off--axis
positions. That consideration supports the possibility of placing the far
detector on surface, thus reducing the experiment cost.

\subsubsection{Systematics in the near--to--far ratio for a set of detector configurations}
\label{subsectsysconf}

\begin{figure}[htbp]
\centering
\includegraphics[scale=0.42,type=pdf,ext=.pdf,read=.pdf]{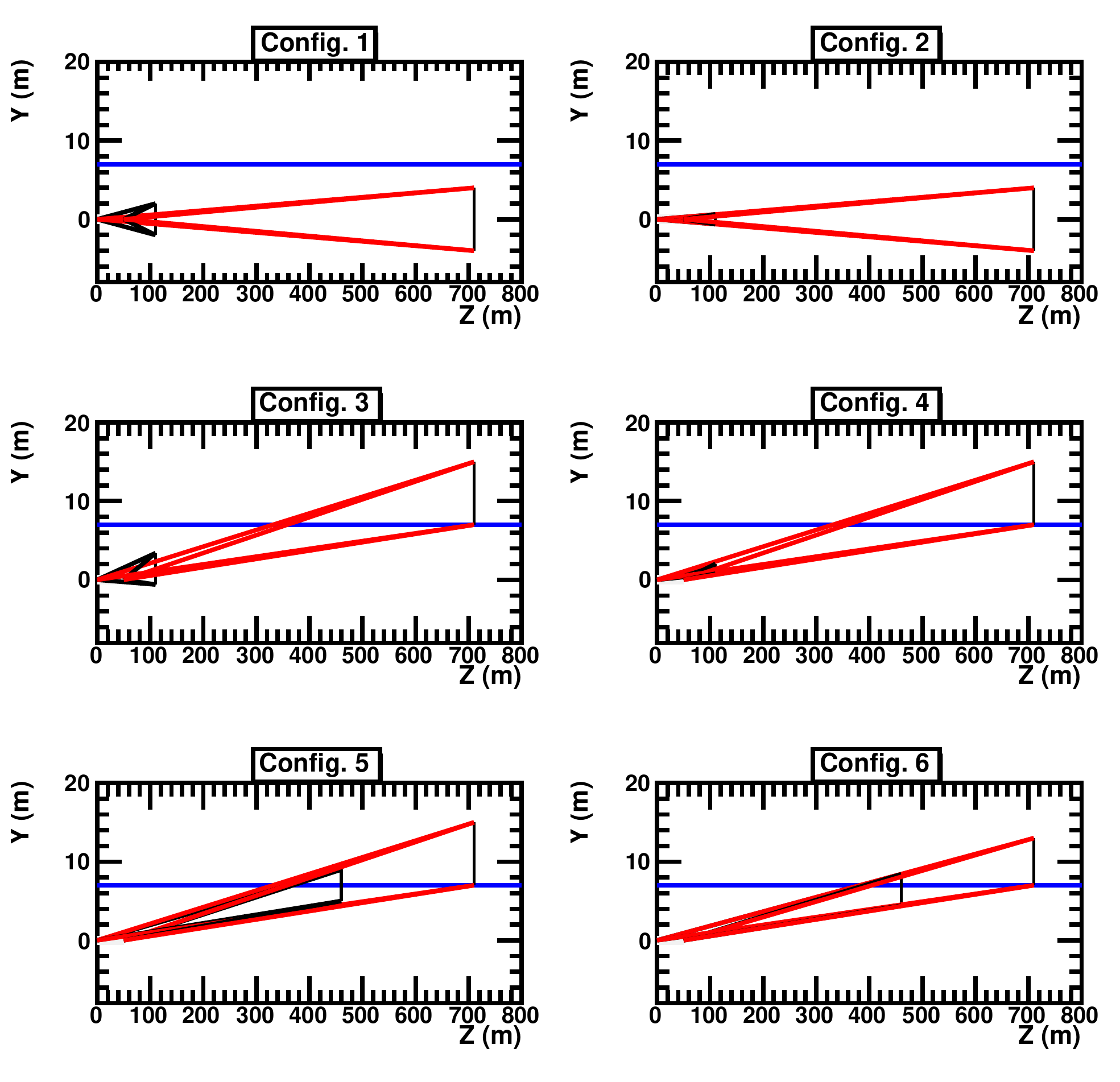}%
\caption{\label{fig:confs}(color online) Configurations of far and near detectors in the $Y$--$Z$ plane (see also Tab.~\ref{tab:confs}).
The blue horizontal line marks the ground level, the vertical black lines mark the detectors and
the red lines show the angle subtended by the detectors at the beginning and the end of the decay pipe.}
\end{figure}

Six configurations were selected considering different 
distances (110, 460 and 710 m), either on--axis or off--axis, and different fiducial sizes of the detectors.  
The configurations' parameters are given in Tab.~\ref{tab:confs} and
illustrated schematically in Fig.~\ref{fig:confs}.


\begin{table}
\caption{\label{tab:confs}Near--Far detectors' configurations. $L_{N(F)}$ is the distance of the near (far) detector
from the target. $y_{N(F)}$ is the vertical coordinate of the center of the near (far) detector with respect to the beam axis,
which lies at about $-$7~m from the ground surface. 
$s_{N(F)}$ is the transverse size of the near (far) detector.}
\scriptsize \centering
\begin{tabular}{ccccccc}
\hline
config. &$L_N$ (m)&$L_F$ (m)&$y_N$ (m)& $y_F$ (m)& $s_N$ (m)& $s_F$ (m)\\
\hline
1 &110&710&0& 0 &4  &8 \\
2 &110&710&0& 0 &1.25  &8 \\
3 &110&710&1.4& 11 & 4 & 8 \\
4 &110&710&1.4& 11 &1.25  &8 \\
5 &460&710&7& 11 & 4 & 8 \\
6 &460&710&6.5& 10 & 4 & 6 \\
\hline
\end{tabular}
\end{table}

\begin{itemize}  
\item Configuration 1 corresponds to two detectors on--axis at 110 (near) and 710 m (far) with squared active
areas of $4 \times 4$ m$^2$ (near) and $8 \times 8~\rm{m}^2$ (far).
By selecting the subsample of neutrinos crossing both the near and far detectors
the region defined in the transverse plane has roughly a squared shape
with a significant ``blurring'' since
 the neutrino source is not point--like. 
\item Configuration 2 make use of a reduced near detector area,
limited to $1.25 \times 1.25$ m$^2$, in order to increase the fraction of neutrinos seen  both at the near and far sites.
\item Configurations 3 and 4 replicate
the same patterns as 1 and 2, respectively, with the far detector on surface and the size of the near
detector defined by the off--axis angle (instead of being both on--axis). 
\item Configurations 5 and 6 are similar to 3 and 4, respectively, but with the near site at a larger distance (460 m).
\end{itemize}



\begin{figure}[htbp]
\centering
\includegraphics[scale=0.45,type=pdf,ext=.pdf,read=.pdf]{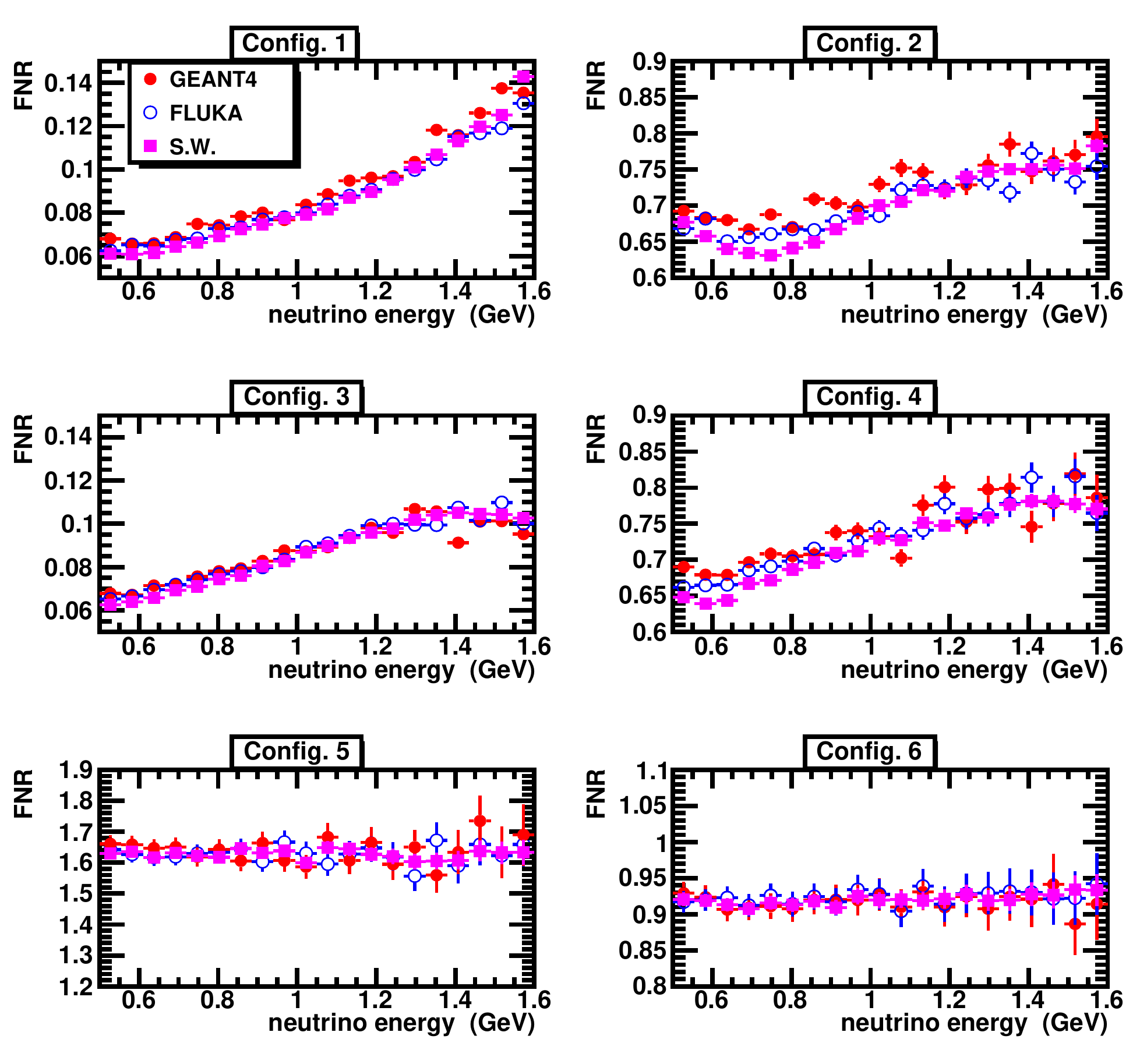}
\caption{\label{fig:fighadp}(color online) Far--to--Near ratios for each considered configuration. Comparison of results from the different
hadro--productions simulations by FLUKA, GEANT4 and the Sanford--Wang parametrization are shown. The error bars indicate only the uncertainty introduced by the limited Monte Carlo samples of FLUKA and GEANT4. The barely visible error bars on the Sanford--Wang
points are due to the very large number of simulated pions.}
\end{figure}

Using FLUKA, GEANT4 or the Sanford--Wang parametrization for the simulation of
$p$--Be interactions, the FNR was computed for each configuration (Fig.~\ref{fig:fighadp}).  

\begin{figure}[htbp]
\centering
\includegraphics[scale=0.42,type=pdf,ext=.pdf,read=.pdf]{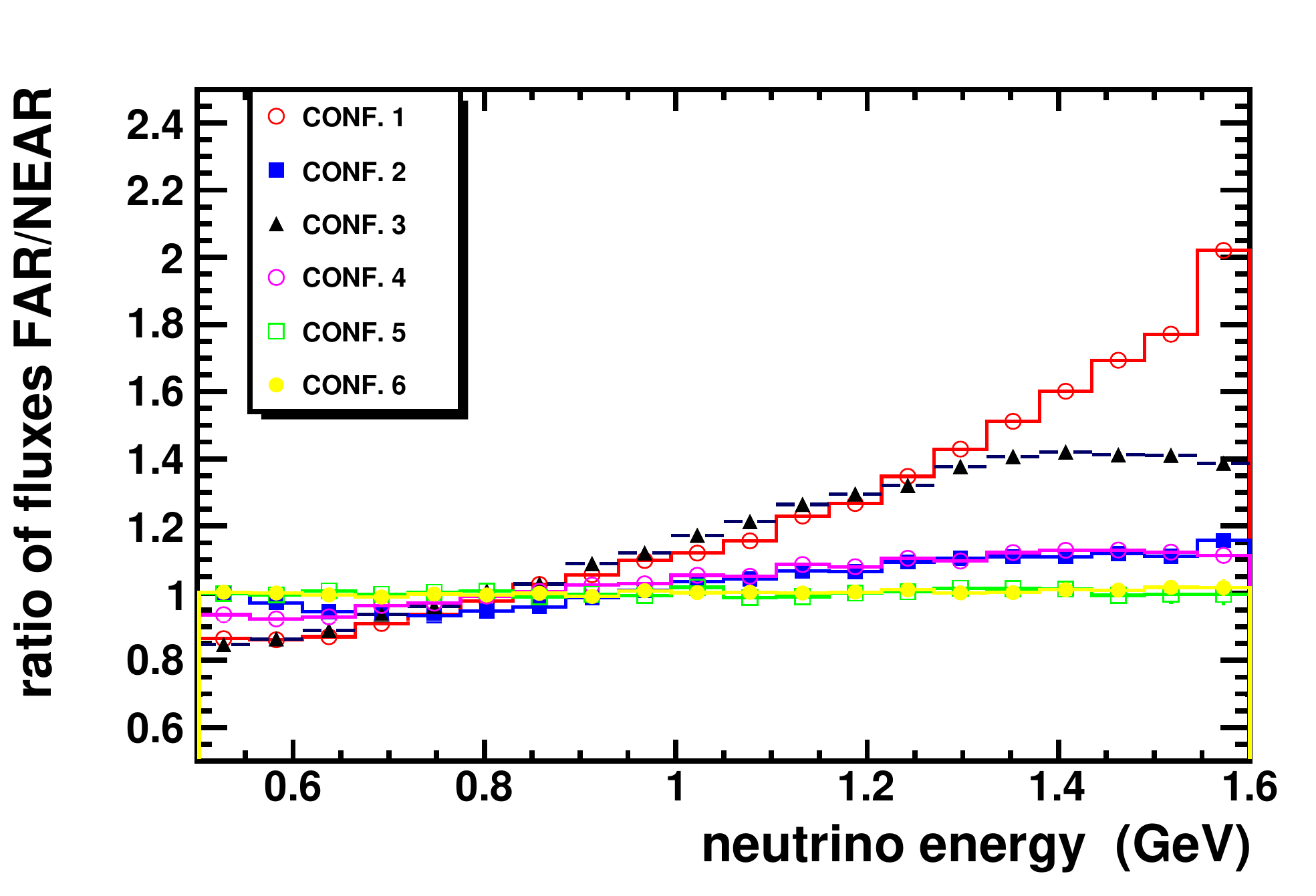}
\caption{\label{fig:fig1}(color online) Far--to--Near ratios for the six considered configurations using the Sanford--Wang parametrization.}
\end{figure}

In configuration 1 (on--axis detectors and a large near--detector size)
the FNR increases with energy, as expected from the discussion in Section~\ref{sec:fnr-ratio}, largely departing from a flat
curve. By reducing the transverse size of the near detector
(configuration 2) the FNR flattens out. This same behavior is
confirmed using off--axis detectors (configurations 3 and
4). Even flatter FNRs are obtained in 
configurations with a near detector at larger baselines (5 and 6).
The different behaviors are more evident in Fig.~\ref{fig:fig1}, where the FNRs based on the  Sanford--Wang parametrization
and normalized to each other are compared.

In order to estimate the impact on the FNR of the hadro--production uncertainties  two studies  were made.

\begin{figure}[htbp]
\centering
\includegraphics[scale=0.45,type=pdf,ext=.pdf,read=.pdf]{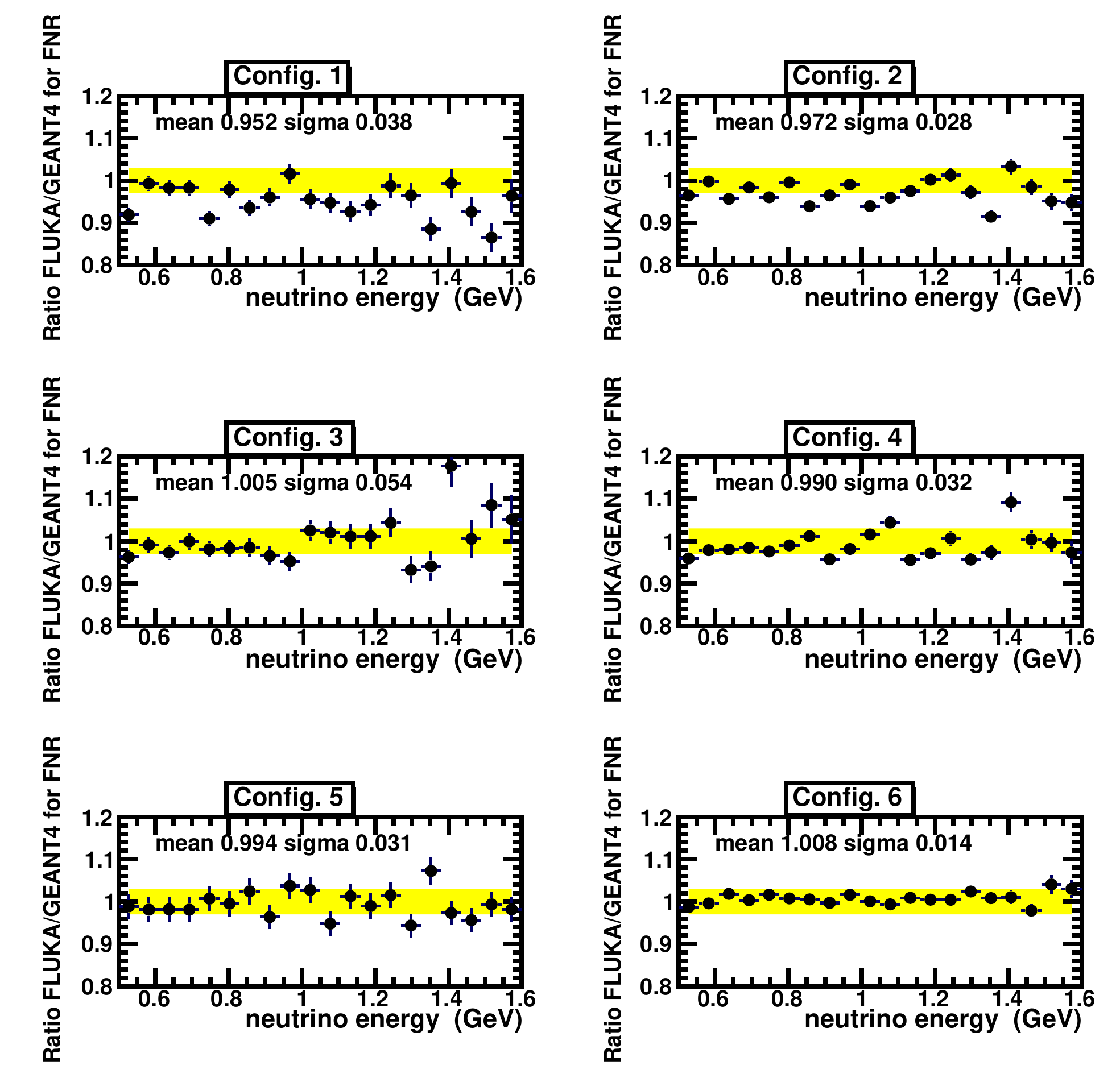}
\caption{\label{fig:fighadp1} Ratio of the two FNRs as predicted by the FLUKA and GEANT4 simulations for 
hadro--production. The bands correspond to a 3\% relative error on the different predictions by FLUKA and GEANT4.}
\end{figure}

First, the difference in the hadronic models implemented in the
FLUKA and GEANT4 generators  were looked at. 
For each configuration the FNR predictions from these two Monte Carlo simulations are drawn in Fig.~\ref{fig:fighadp1} as their ratio.
The (yellow) bands correspond to a fixed 3\% error on the FNR ratios between FLUKA and GEANT4. 
The two simulations agree at 1 to 3\% level when 
an overlapping region between the far and near detectors occurs (configurations 2, 4 and 6).


%



\begin{figure*}[htbp]
\centering
\includegraphics[scale=0.45,type=pdf,ext=.pdf,read=.pdf]{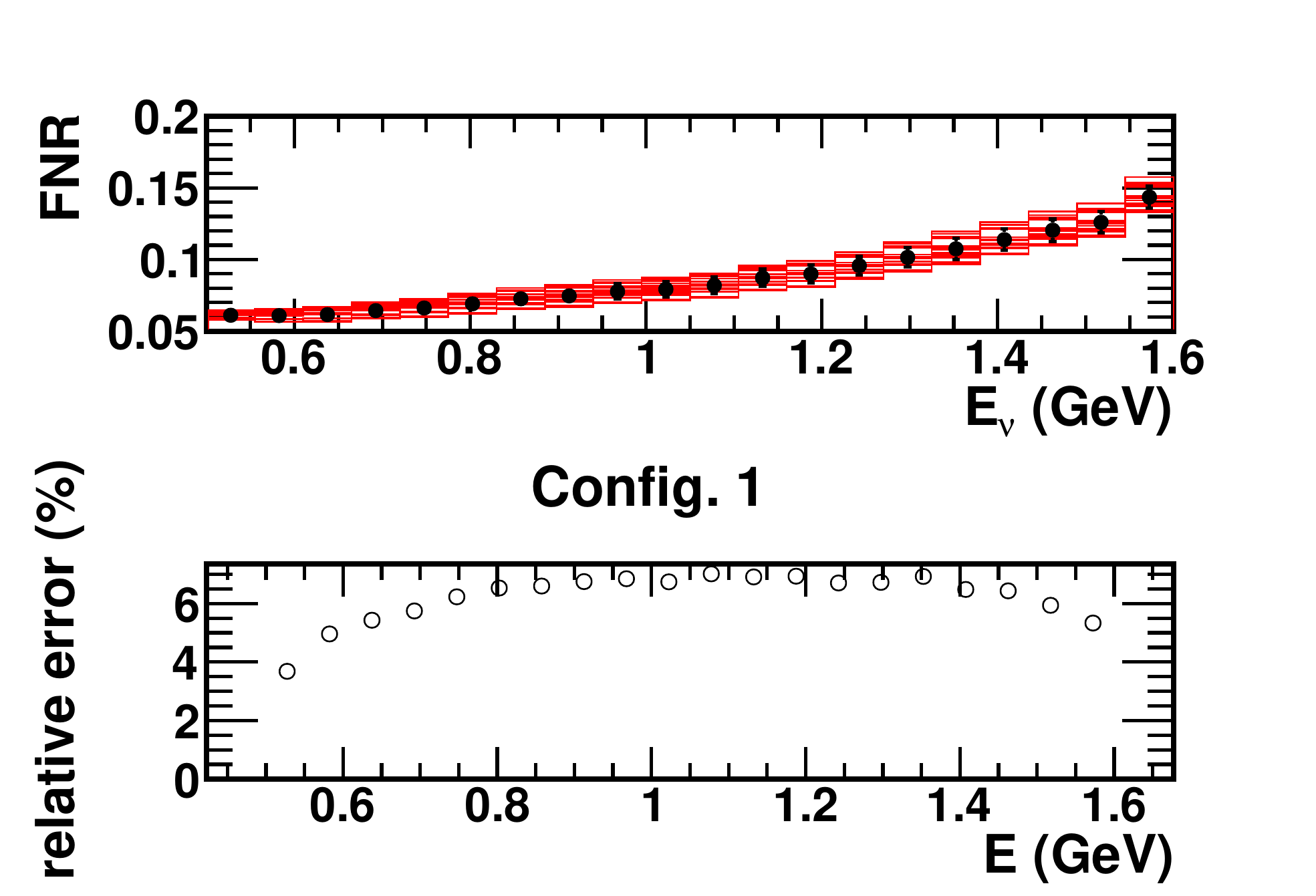}%
\includegraphics[scale=0.45,type=pdf,ext=.pdf,read=.pdf]{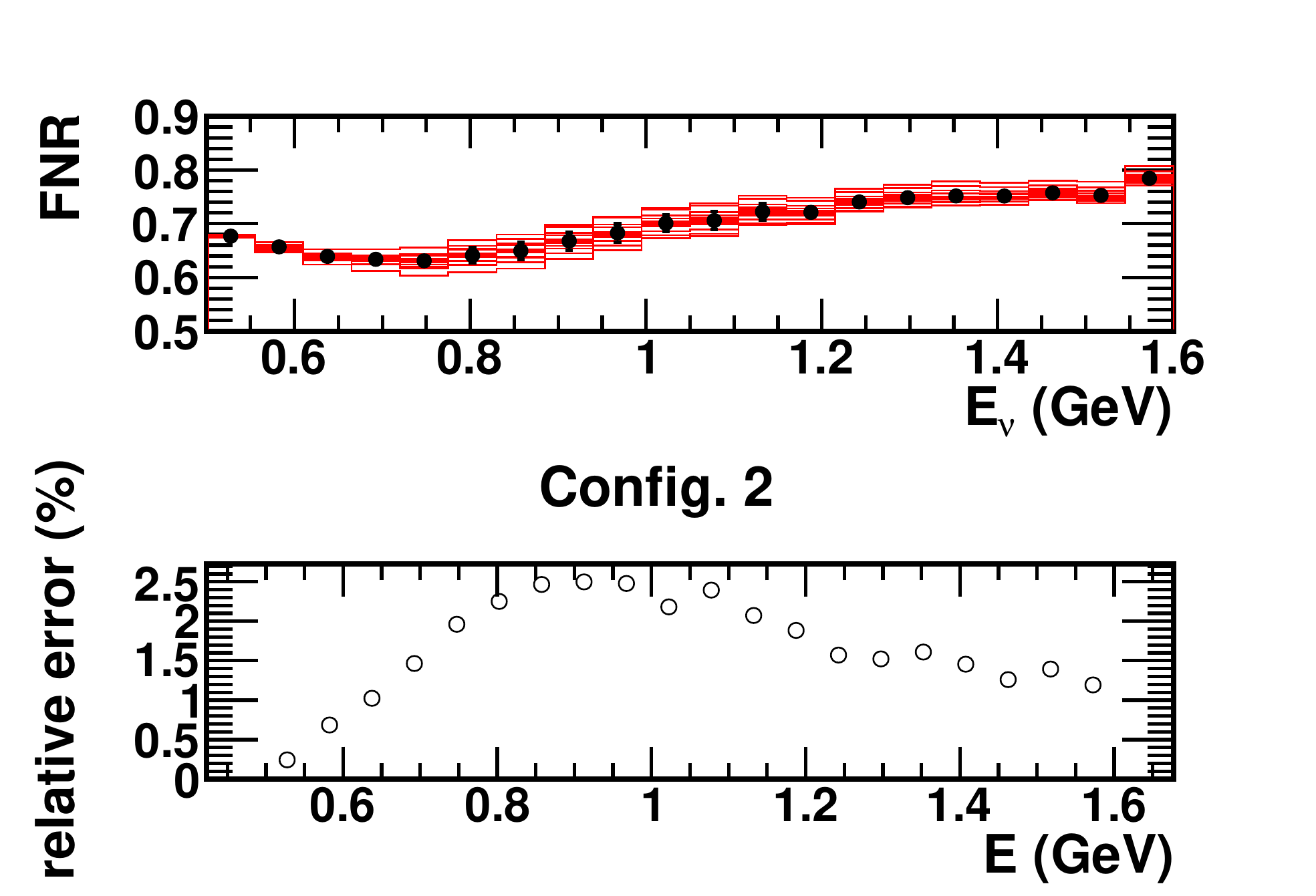}
\includegraphics[scale=0.45,type=pdf,ext=.pdf,read=.pdf]{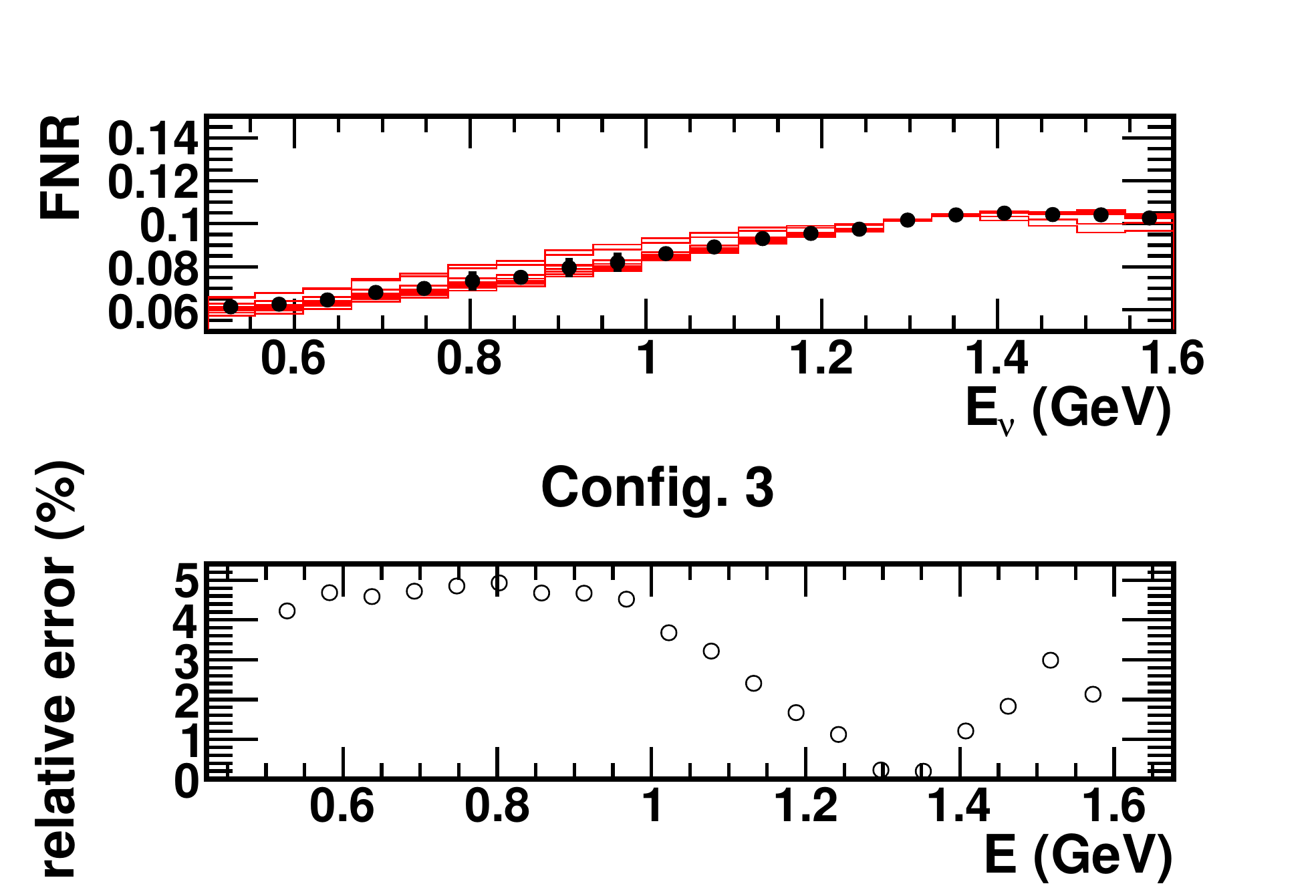}%
\includegraphics[scale=0.45,type=pdf,ext=.pdf,read=.pdf]{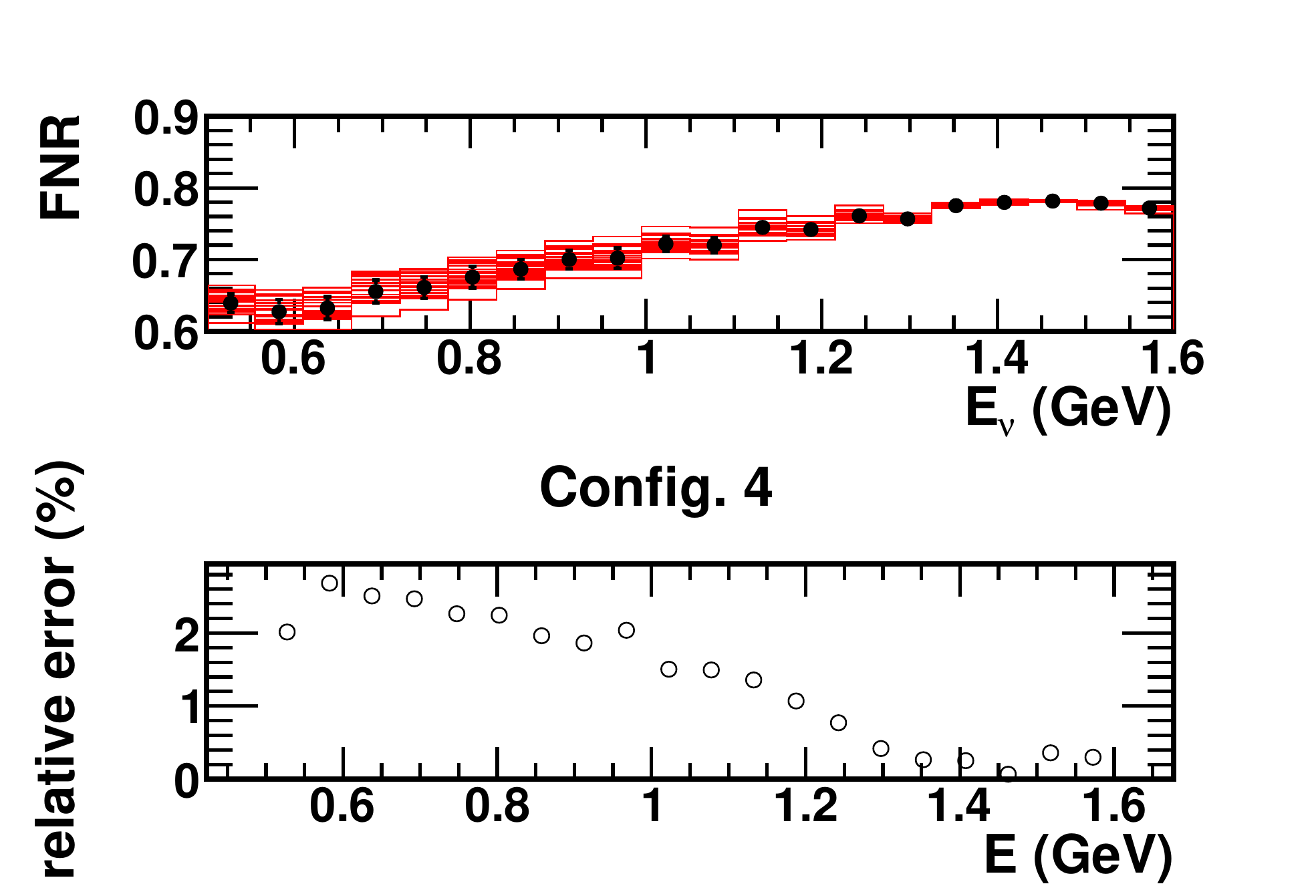}
\includegraphics[scale=0.45,type=pdf,ext=.pdf,read=.pdf]{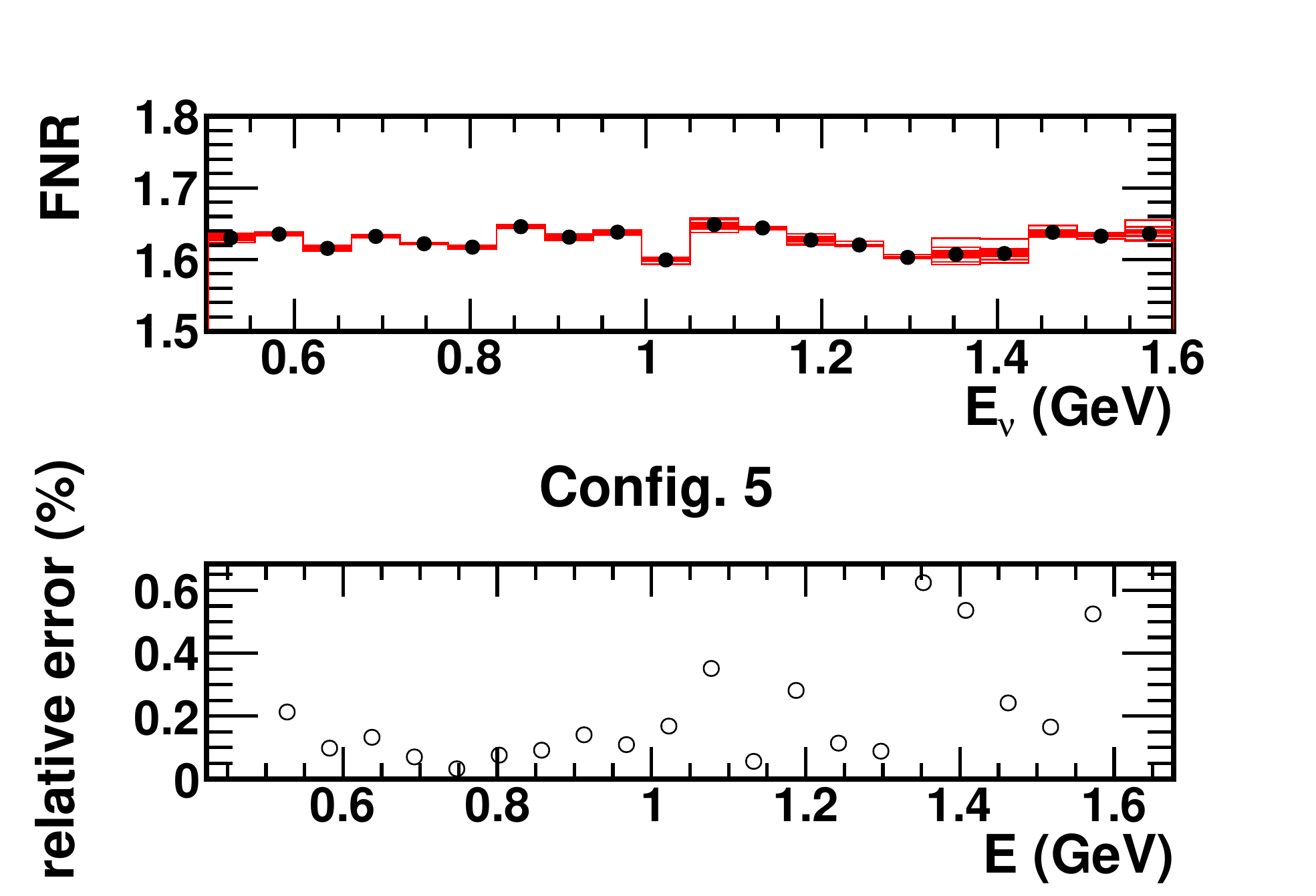}%
\includegraphics[scale=0.45,type=pdf,ext=.pdf,read=.pdf]{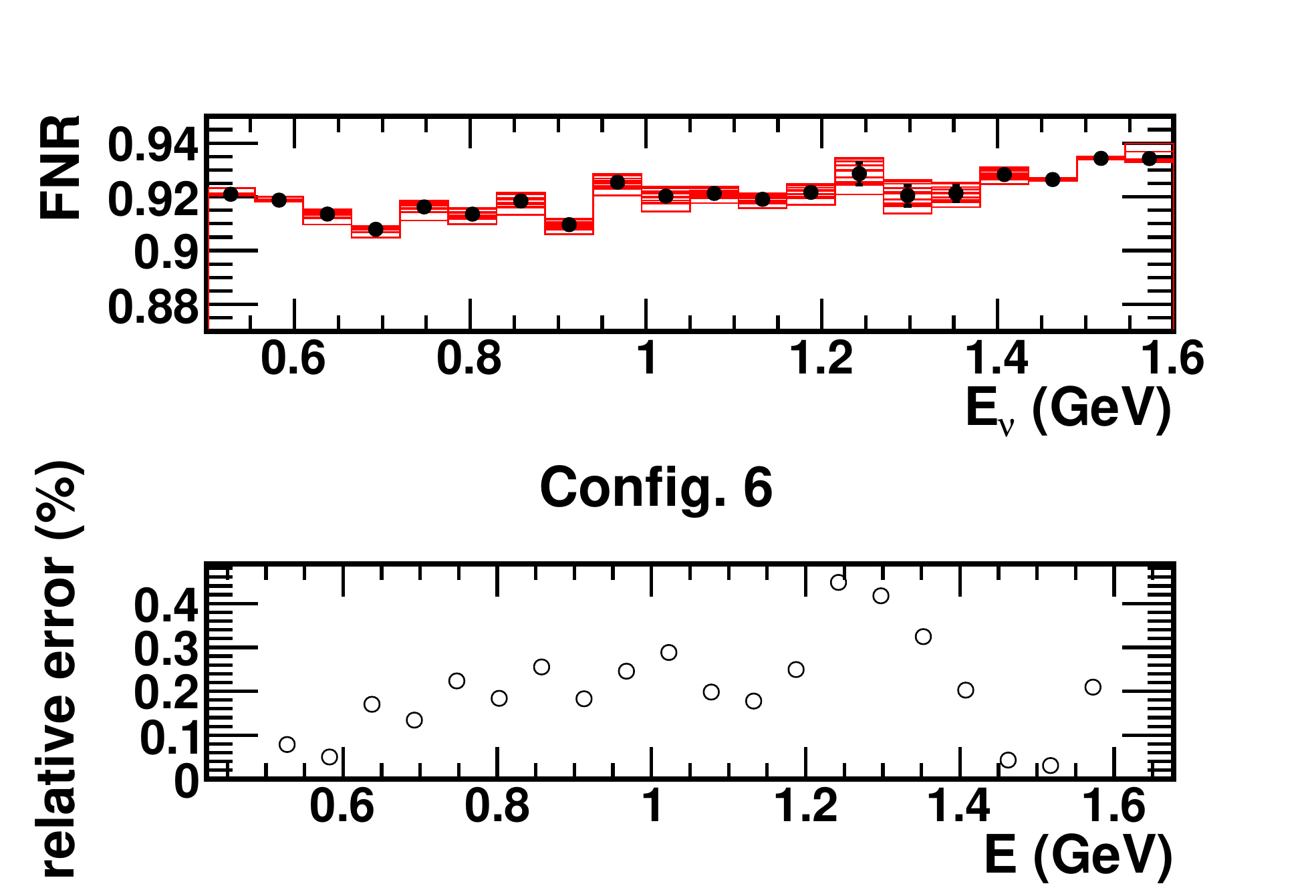}
\caption{\label{fig:SWrew}The FNR results for the six configurations are shown as function of the neutrino energy.
The effect of data--driven hadro--production uncertainties on the FNR computed using the 
Sanford--Wang parametrization for the six configurations is depicted. Histogram lines
show typical individual--samplings of the $c_i$ parameters of Eq.~\ref{eq:SW}. 
Plain bullets correspond to the
average value in each bin while the error bars represent the r.m.s. of the samplings.
For each configuration, bottom plots (hollow bullets) show the ratios of the r.m.s. over the central values,
 providing an estimate of the fractional systematic error.}
\end{figure*}

Another approach was adopted to investigate the FNR syste\-ma\-tic--error due to hadro--production based on the
existing measurements and the corresponding covariance--error matrix of Booster Be--target replica~\cite{G4BNBflux}.
The coefficients $c_i$ of the Sanford--Wang parametrization of
pion production data from HARP and E910 in Eq.~\ref{eq:SW} were
sampled within their correlation errors. 
%
%
The sampling of these correlated parameters was performed via
the Cholesky decomposition of the covariance matrix
reported in Tab.~5 of~\cite{G4BNBflux}. 
%
For each sampling of the $c_i$ coefficients, neutrinos were weighted with a factor
\begin{equation}
w(p_\pi, \theta_\pi)=\frac{\frac{d\sigma}{dp_\pi d\theta_\pi}(c_i)}{\frac{d\sigma}{dp_\pi d\theta_\pi}(c^0_i)}
\end{equation}
depending on the momentum ($p_\pi$) and angle ($\theta_\pi$) of their
parent pion. $c^0_i$ are the best--fit values to the
HARP--E910 data--set. The FNR  for different $c_i$ varied within their covariance error--matrices
are shown in Fig.~\ref{fig:SWrew} for the six  considered configurations. 

\noindent For each configuration, in the top plots (plain bullets) the average value is shown with its error bar representing 
the r.m.s. of the samplings. Bottom plots (hollow bullets) show the ratio of the r.m.s. 
over the central value providing an estimate of the fractional systematic error.
Uncertainties are rather large (5--7\%) when considering the full area of the near detector at 110~m;
they decrease significantly when restricting to the central region (configurations 2, 4 and 6). In particular, in configuration 4,
that is a realistic one from practical considerations, the uncertainty ranges from 2\% at
low neutrino energy and decreases below 0.5\% --1.5\% at neutrino energies above 1~GeV.
The uncertainty is generally below 0.5\% for a near site at 460~m.

\subsection{Conclusions for Section~\ref{sec:beam}}
Full simulations of the Booster beam were made anew, based on FLUKA and GEANT4 Monte Carlo simulations. 
Indications of the  systematic error on the far--to--near ratio were obtained, showing characteristic behaviours. 
Moreover,
using the constraints from HARP--E910 data  the uncertainties on the FNR, associated to hadro--production, were 
carefully estimated.

Six configurations of the detector locations and sizes were considered. 
For a far detector on ground--surface and a near detector at the same off-axis angle the systematic error stays at 1--2\%
when the far and near transverse surfaces are matched in acceptance (configuration 4).
Provided the high available statistics and the large lever--arm for oscillation studies,
the layout with baselines of 110~m and 710~m is considered in the following as the best choice.

\section{Detector Design Studies}\label{sec:spect1}

The location of the Near and Far sites is a fundamental issue in a search for sterile neutrino at SBL. Moreover the two detector systems 
at the two sites have to be as similar as possible. The NESSiE far and near spectrometer system were designed to 
match with a timely
schedule and also exploit the  experience acquired in the construction, assembling and 
maintenance of the OPERA spectrometers~\cite{bopera} that own an active transverse area of $8.75\times 8.00$ m$^2$ . 

The OPERA two large dipole iron magnets will be dismantled in 2015--2016   
and possibly be re-used for the \numu disappearance study discussed in this paper.
They are made of two vertical arms connected by a top and a bottom return yoke. 
Each arm is composed of 12 planes of  5 cm thick iron slabs, interleaved by 11 planes of Resistive Plate Chambers (RPC)  that provide the inner tracker. 
The magnetic field has opposite directions in the two arms and is uniform in the tracking region, with an intensity of 1.53~T. 

Muons stopping in the spectrometers can be identified by their range. Their fraction can be maximized by increasing the depth of the magnets. This can be achieved by longitudinally coupling the two OPERA spectrometers, both at the far and the near sites, minimizing therefore the detector re--design.
Their modularity allows to cut every single piece at 4/7 of its height, using the bottom part for the far site and the top part
for near site. In this way  any inaccuracy in geometry (the single 5 cm thick iron slab owns a precision of few mm) or any variation of 
 the material properties with respect to the nominal ones (they are at the level of few percent) will be the same in the two detection sites.
The near NESSiE spectrometer will thus be a clone of the far one, with identical thickness along the beam but a reduced 
transverse size. 
 
\begin{figure}[htbp]
\begin{center}
\includegraphics[scale=0.4]{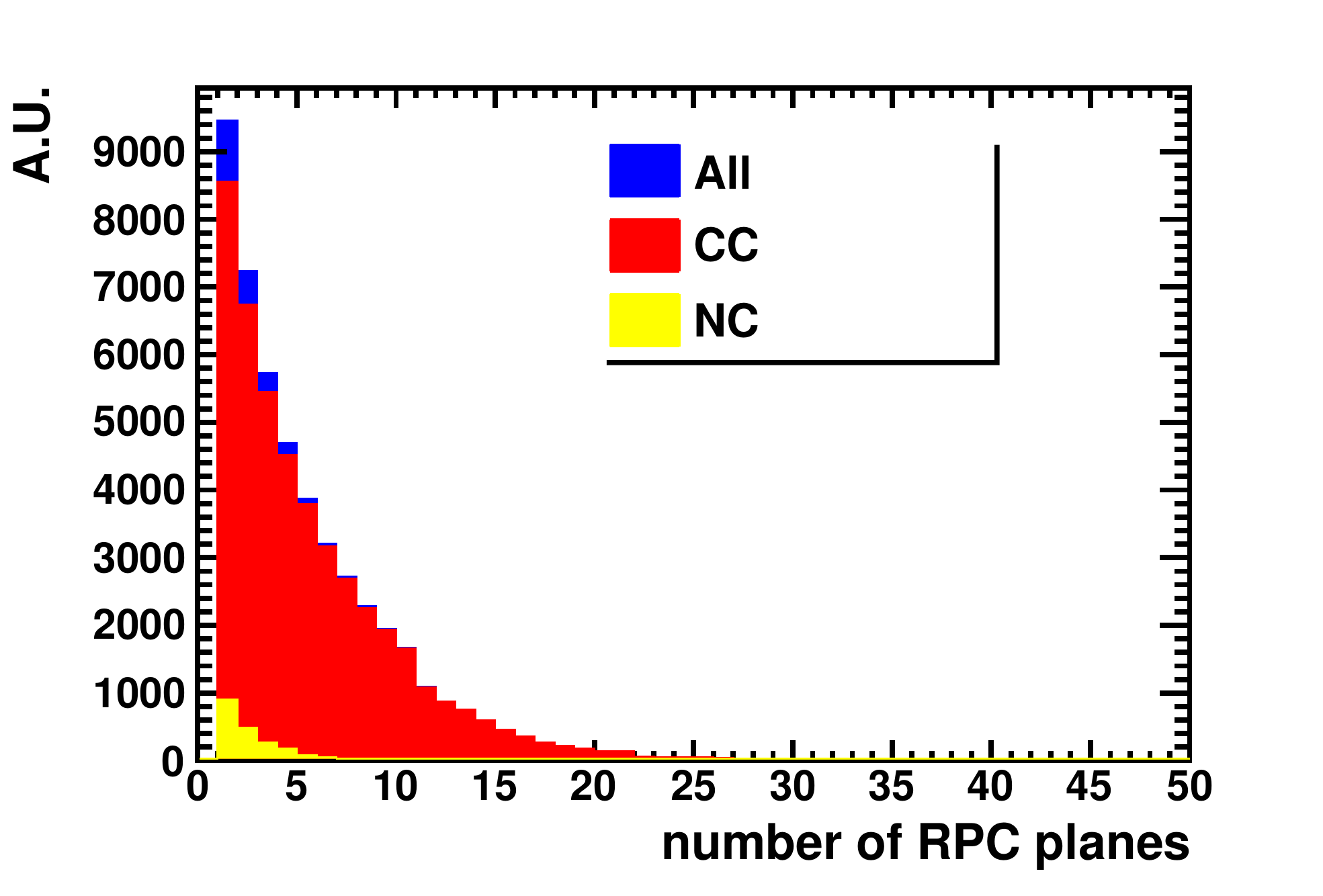}
\caption{\label{fig:RPCplanes_5cm}(color online) Number of crossed RPC planes for  charged and neutral current events (5 cm slab geometry). }
\end{center}
\end{figure}

With the proposed  setup a very large fraction of muons from CC neutrino interactions is stopped in the spectrometer. 
For this class of events the momentum is obtained by muon range. For higher energies the muon momentum is determined from the muon track sagitta measured in the bending plane. The charge of the muon can be determined when its track crosses few RPC planes ($\ge$ 3).
The distributions of hit RPC planes are shown in Fig.~\ref{fig:RPCplanes_5cm} for charged and neutral current events.

\begin{figure}[htbp]
\begin{center}
\includegraphics[scale=0.43]{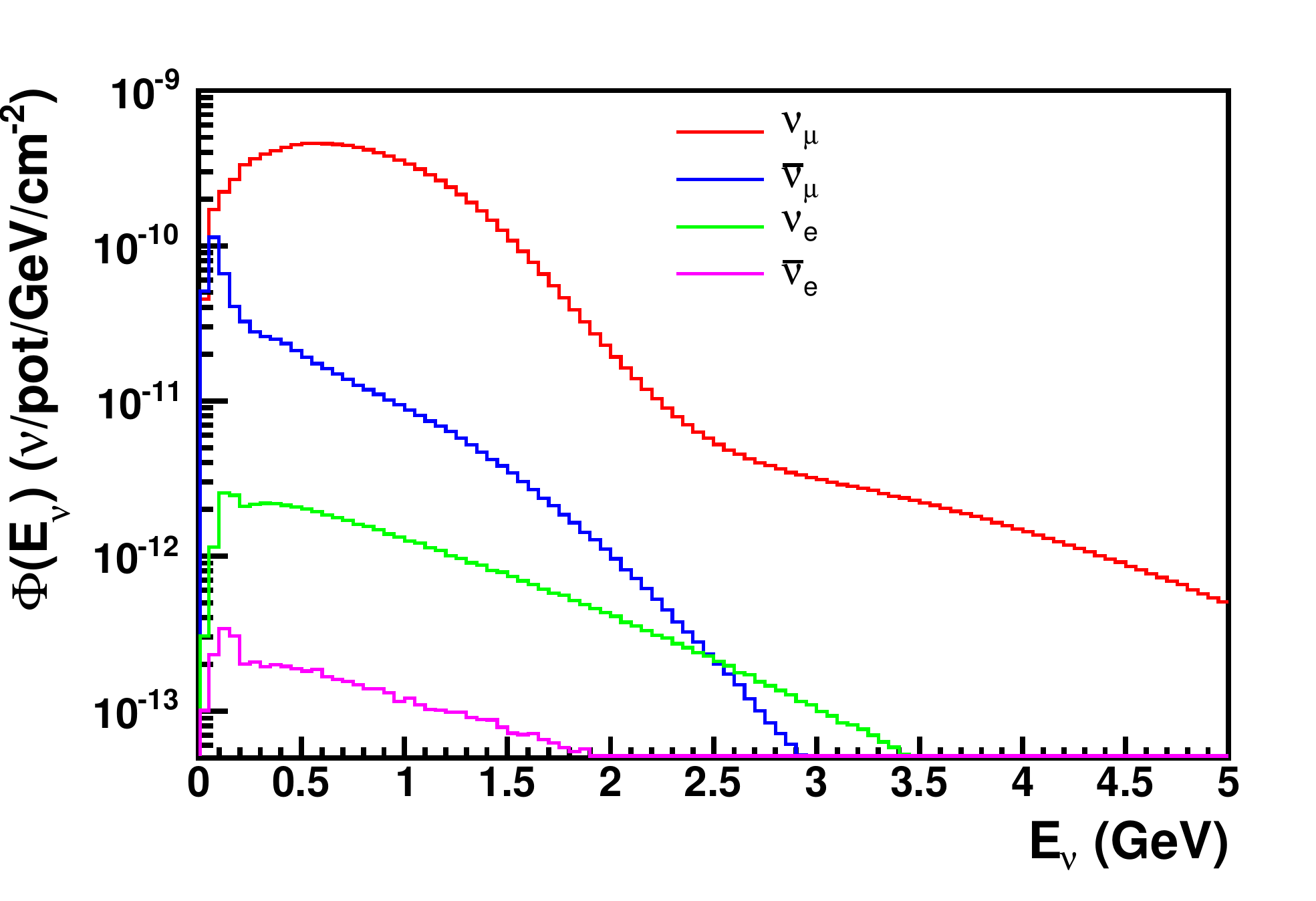}
\caption{\label{fig:booster-pos}(color online) The fluxes of the Booster neutrino beam (from~\cite{G4BNBflux}).}
\end{center}
\end{figure}

In the positive--mode running of the Booster beam the antineutrino contamination is rather low (see Fig.~\ref{fig:booster-pos}).
In this case the use of the charge identification is limited and it can contribute
only to keep the related systematic error under control and well below 1\% since the mis--identification of the charge
(mis---ID) of the Spectrometers is about 2.5\%
in the relevant momentum range (see the bottom plot of Fig.~\ref{fig:norma-interac-anti}). The situation is quite different for the 
negative--mode running where the neutrino contamination is rather high (see Section~\ref{sec:antinu}).

In the following the performances (efficiency and purity) of the spectrometer with 5 cm thick iron--slabs are evaluated in terms of Neutral Current (NC) contamination and momentum resolution, and compared to a geometry with 2.5 cm thick iron slabs. 

With 5 cm slabs the fraction $\varepsilon$ of neutrino interactions in iron inducing at least one RPC hit 
($\varepsilon \equiv {(\geq 1RPC)}/{all}$) is 68\%. 
The efficiency 
for CC and NC  events is $\varepsilon_{CC} = 86\%$ and $\varepsilon_{NC} = 20\%$, respectively. 
That corresponds to a fraction of NC interactions over the total number of interaction ${NC}/{all} = 8.1\%$. 
With a minimal cut of 2 crossed RPC planes, the NC contamination is reduced to 4.2\%; by requiring 3 RPC planes the NC contamination 
drops to 3.0\%. 

\begin{figure}[htbp]
\begin{center}
\includegraphics[scale=0.4]{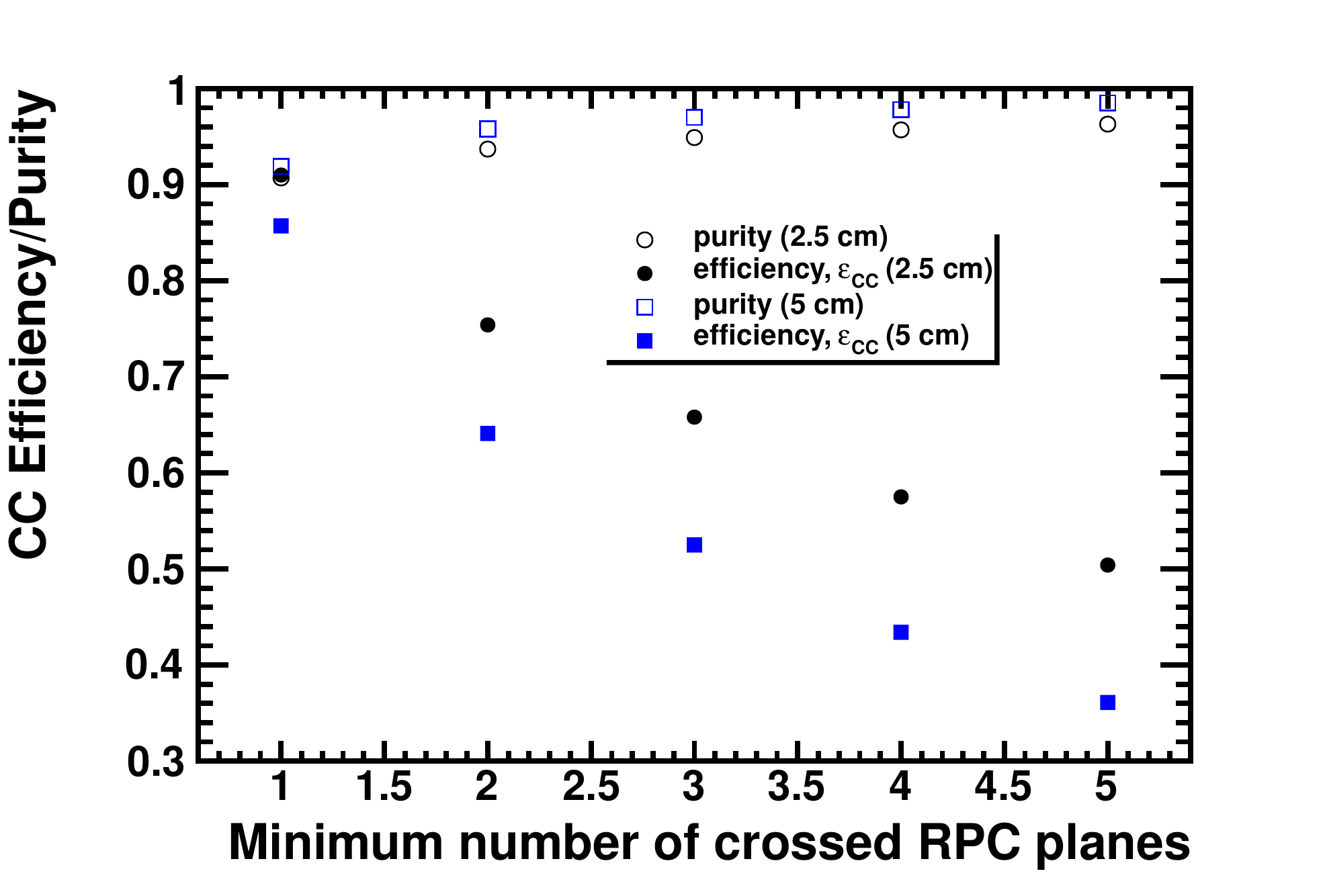}
\caption{\label{fig:eff_pur}CC efficiency ($\varepsilon_{CC}$) and purity as a function of the minimum number of hit
RPC planes for the two spectrometer geometries, 5 cm thick iron slabs (squares) and 2.5 cm thick slabs (circles).}
\end{center}
\end{figure}

Using 2.5 cm thick slabs, the fraction of neutrino events with at least one RPC hit  increases for both NC and CC events. 
The CC efficiency and the NC contamination are both larger with respect to the 5 cm geometry. 
In Fig.~\ref{fig:eff_pur}  $\varepsilon_{CC}$ and the purity, $CC/all = 1 - NC/all$, are shown 
as a function of the minimum number of crossed RPC planes, for either slab thickness. 

At the same level of purity the efficiencies in the two geometries are comparable. 
No advantage in statistics is obtained with thinner slabs if the same NC contamination suppression is required. 
In conclusion the already available 5 cm thick iron slabs can be adopted. 
It is worthwhile to note that purities and efficiencies have been extensively checked not to be spoiled by second-order systematic effects 
due to the convolution of the fluxes and the neutrino cross-section with the detector acceptances at the two sites. 
That keeps the systematic error due to the neutrino detection below 1\% while the most relevant contribution to systematics 
remains the uncertainty on the fluxes\footnote{More discussion is provided in the proposal~\cite{nessie-fnal}.}.


\subsection{Track and momentum reconstruction}
The RPC digital read--out is provided on both vertical ($Y$) and horizontal ($X$) coordinates using 2.6 cm pitch strips. 
Track reconstruction is made first in the two RPC projections  (the $XZ$ bending plane, and the $YZ$ plane). Then,
 the two 2--D tracks are merged to reconstruct a 3--D event. 

\begin{figure}[htbp]
\begin{center}
\includegraphics[scale=0.4]{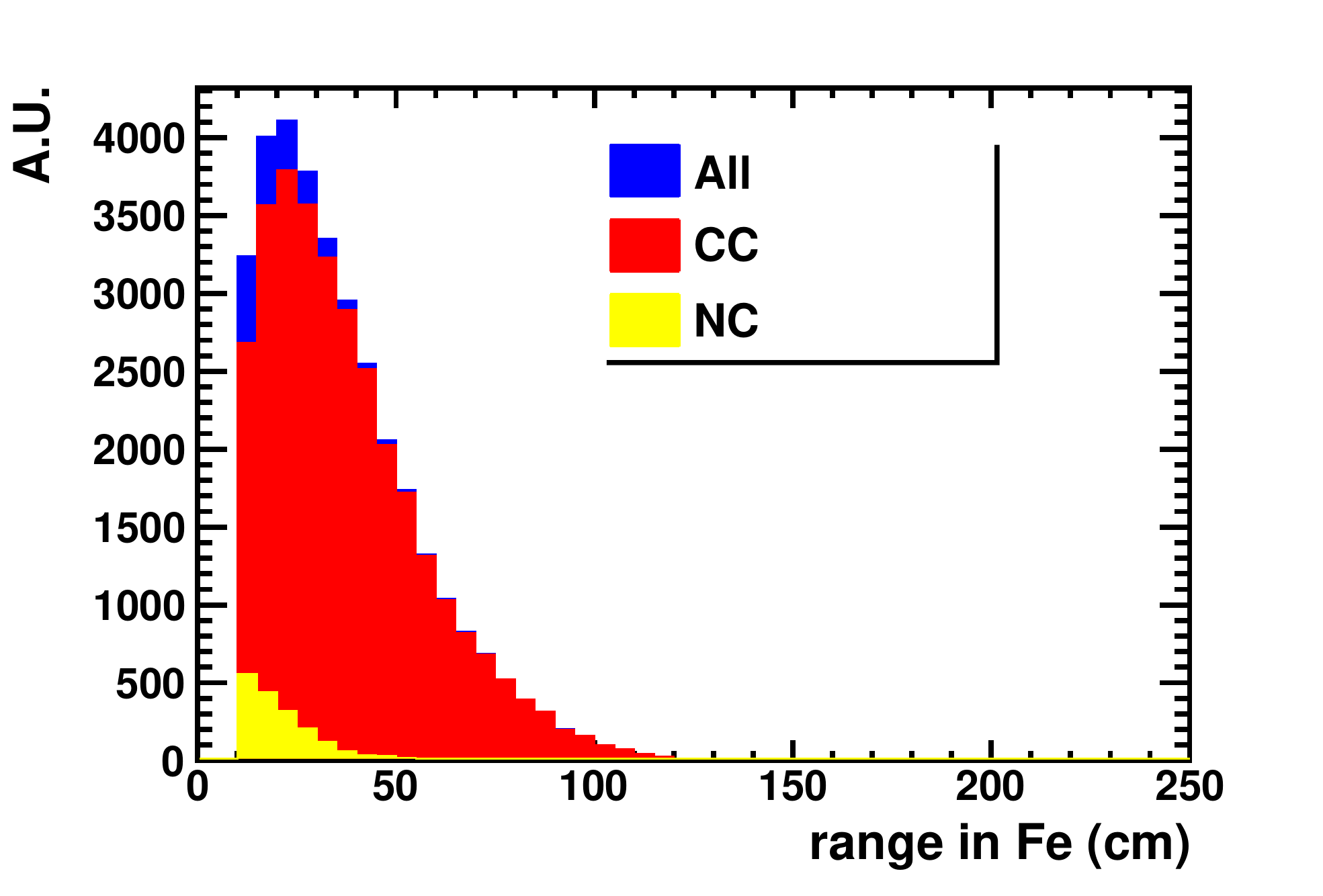}
\caption{\label{fig:range}(color online) Muon range of reconstructed events, for CC plus NC (blue), CC (red) and NC (yellow) events. }
\end{center}
\end{figure}

For muons stopping inside the spectrometer  
 the  momentum is obtained from the track range using the con\-ti\-nu\-ous--slowing--down approximation~\cite{pdg}. 
The range distribution is shown in Fig.~\ref{fig:range}. Similar conclusions can be drawn as for the number of crossed RPC planes
(in Fig.~\ref{fig:RPCplanes_5cm}), namely the CC efficiency and the reduction of the NC related background.
For any muon track a parabolic fit is performed in the bending plane to evaluate the track sagitta thus determining particle charge and momentum. 

\begin{figure}[htbp]
\begin{center}
\includegraphics[scale=0.4]{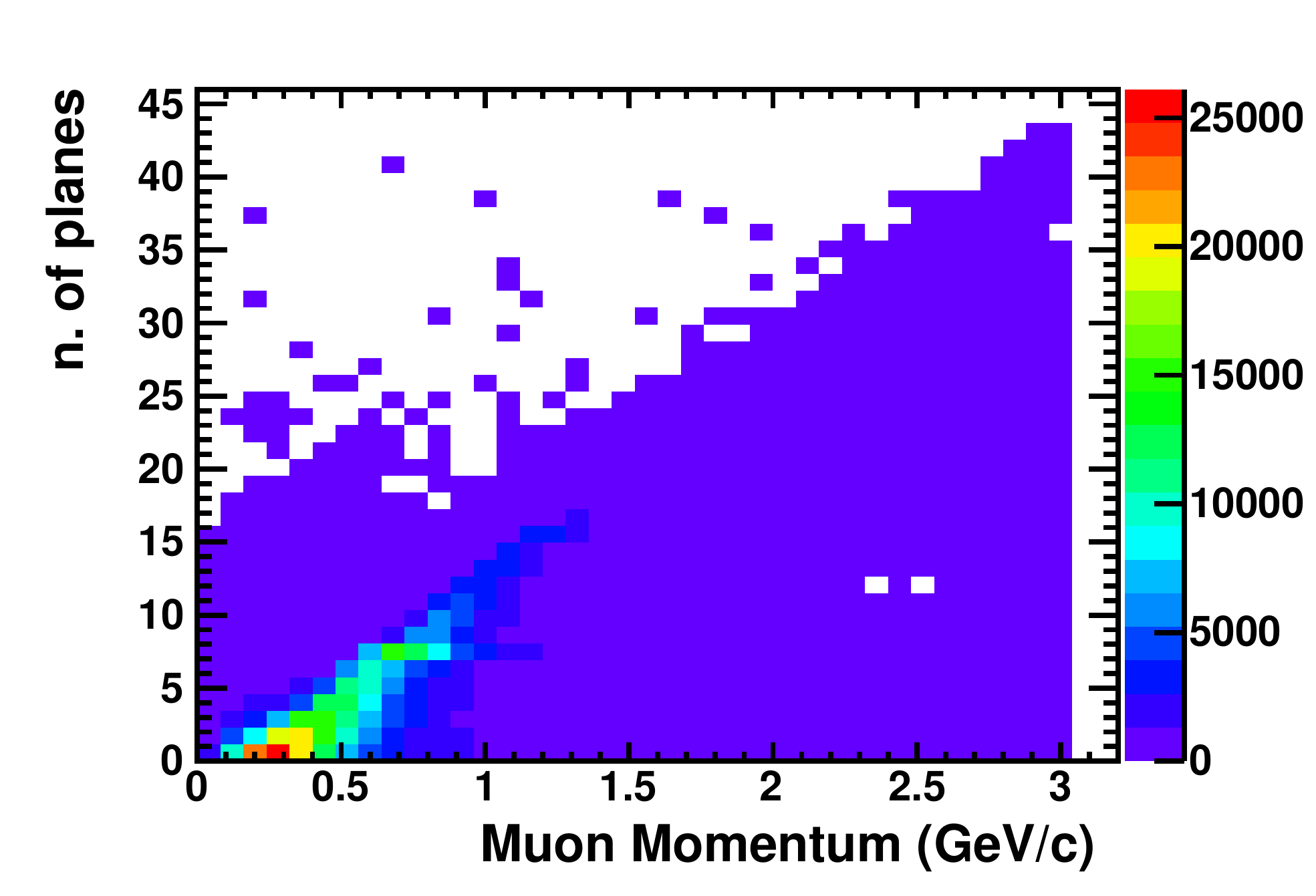}
\includegraphics[scale=0.4]{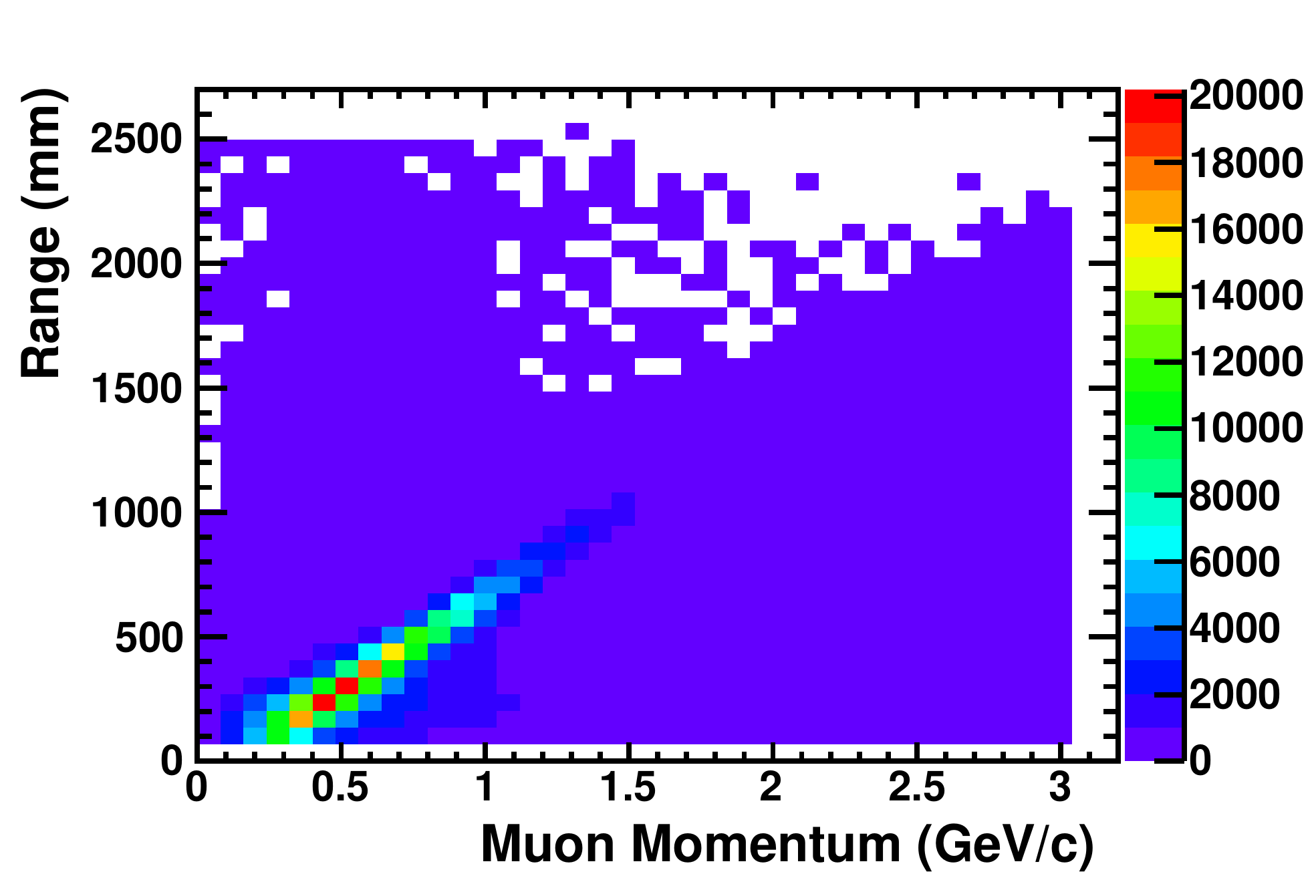}
\caption{\label{fig:smearing}(color online) Number of crossed RPC planes (top) and range (bottom) versus muon momentum (CC events). }
\end{center}
\end{figure}

In Fig.~\ref{fig:smearing} (CC events) the reconstructed variables, namely the number of fired RPC planes and the range in iron, are plotted versus the 
muon momentum. 
A correlation is visible for both variables. The very strong muon 
momentum--range linear correlation  allows to reach a sensitivity of few 
percent in the momentum estimation.


\subsection{Conclusions for Section~\ref{sec:spect1}}
In the previous Section~\ref{sec:beam} it was shown that by adopting a realistic layout configuration, and by a proper choice of the 
fiducial volume at the near site, a reduction of the uncertainty on the Far/Near ratio to less than 2\% level is possible,
by plugging the data--driven knowledge on hadro--production (HARP and E910~\cite{G4BNBflux}).  This is by far the 
dominant component. 
Possible other effects due to the running conditions 
of the detectors once installed can be kept under control ($<1\%$).
It must be noted that the efficiency and acceptance of the spectrometers can be checked 
routinely with good accuracy using the large amount of available cosmic rays. Furthermore the detector in itself is very well 
mastered and understood due to its simplicity and to the extensive experience in running it underground on the CNGS beam.
Furthermore, each original (from OPERA) iron slab will be used partly at the near and partly at the far site, providing the same 
geometrical and material composition. 
That choice would provide a very constrained system at the two sites, with not only identical targets, but also similar geometrical frames and acceptances. The relative large statistical sample that could be obtained within configuration 4 would allow a careful control of the related systematic effects, by operating at different energy ranges, too. Finally, all the effects due to detector acceptance and event--reconstruction have been evaluated to be within 1\%. 

\section{Physics Analyses and Performances}\label{sec:analysis}

The disappearance probability of muon--neutrinos, $P(\nu_\mu\to {\rm not}\, \nu_\mu)$, in presence of an additional sterile--state 
can be expressed in terms of the extended PMNS~\cite{pmns} mixing matrix ($U_{\alpha i}$ with $\alpha = e, \mu, \tau, s$, and $i = 1,\ldots,4$). 
In this model, called ``3+1'', the neutrino mass eigenstates $\nu_1,\ldots,\nu_4$ are labeled such that the first three states are mostly made 
of active flavour states and contribute to the ``standard'' three flavour oscillations with the squared mass differences 
$\Delta m_{21}^2 \sim 7.5\times 10^{-5}~{\rm eV^2}$ and $|\Delta m_{31}^2| \sim 2.4\times 10^{-3}~{\rm eV^2}$, 
where $\Delta m_{ij}^2 = m^2_i - m^2_j$. The fourth mass eigenstate, which is mostly sterile, is assumed to be much heavier than the others, 
$0.1~{\rm eV^2}\lesssim \Delta m_{41}^2 \lesssim 10~{\rm eV^2}.$ The opposite case in hierarchy, i.e. negative values of $\Delta m_{41}^2$, produces a similar phenomenology from the 
oscillation point of view but is disfavored by cosmological results on the sum of neutrino masses~\cite{cosmo-data}. 

In a Short-Baseline experiment the oscillation effects due to $\Delta m^2_{21}$ and $\Delta m^2_{31}$ can be neglected since $L/E\sim 1$ km/GeV. 
Therefore the oscillation probability depends only on $\Delta m^2_{41}$ and $U_{\alpha 4}$ with $\alpha = e,\mu,\tau$.
In particular the survival probability of muon neutrinos 
is given by the effective two--flavour oscillation formula:
\begin{equation}
P(\nu_{\mu}\to\nu_{\mu})_{SBL}^{3+1} = 1 - \left[ 4 \vert U_{\mu 4} \vert^2 (1 - \vert U_{\mu 4} \vert^2)\right]  \sin^2 \frac{\Delta m^2_{41} L}{4E},
\end{equation}
where 
$4 \vert U_{\mu 4} \vert^2 (1 - \vert U_{\mu 4} \vert^2)$ is the {\em amplitude} and, since the baseline $L$ is fixed by the experiment location, 
the oscillation {\em phase} is driven by the neutrino energy E.

In contrast, appearance channels (i.e. $\nu_\mu \to \nu_e$) are driven by
terms that mix up the couplings between the initial and final flavour--states and
the sterile state, yielding a more complex picture:
\begin{equation}
P(\nu_{\mu}\to\nu_e)_{SBL}^{3+1} = 4 \vert U_{\mu 4}\vert^2 \vert U_{e 4} \vert^2  \sin^2 \frac{\Delta m^2_{41} L}{4E}
\end{equation}
Similar formulas hold also assuming more sterile neutrinos ($3 + n$ models).

Since $\vert U_{\alpha 4}\vert$ is expected to be small, the appearance channel is suppressed by two more
powers in $\vert U_{\alpha 4}\vert$ with respect to the disappearance one. Furthermore, since $\nu_e$ or $\nu_\mu$ appearance
requires $\vert U_{e 4}\vert > 0$ and $\vert U_{\mu 4}\vert > 0$, it should be naturally accompanied by
non--zero $\nu_e$ and $\nu_\mu$ disappearances. In this sense the disappearance
searches are essential for providing severe constraints  on the theoretical models 
(a more extensive discussion on this issue can be found e.g. in Section~2 of~\cite{winter}).

It should also be noticed that a good control of the $\nue$ contamination is
important when using the $\nu_\mu \to \nu_e$ for sterile neutrino searches at SBL.
In fact the observed number of  $\nu_e$ neutrinos would depend on the $\nu_\mu\rightarrow\nu_e$ appearance
and also on the  $\nu_e\rightarrow\nu_s$ disappearance. 
On the other hand, the amount of $\nu_\mu$ neutrinos would be affected  by the
$\nu_\mu\rightarrow\nu_s$  and $\nu_e\rightarrow\nu_\mu$ transitions. However the latter term (\numu appearance)  
 would be much smaller than in the \nue case since the $\nu_e$ contamination in $\nu_\mu$ beams is
usually at the percent level. 
In conclusion in the $\nu_{\mu}$ disappearance channel the oscillation probabilities in either the near or far detector 
are not affected by any interplay of different flavours. Since both near and far detectors measure the same single
disappearance transition, the probability amplitude is the same at both sites.

Another important aspect of the analysis is related to the procedure of either rejecting or evaluating the presence of 
a sterile component. The basic hypothesis of no-sterile oscillation ($H_0$) has been assumed against the presence of 
something else ($H_1$). For $H_0$ the analysis is simplified since the systematic errors on the FNR estimator 
are definitively under control when no-sterile component is included, as illustrated in the previous sections. In particular the cross-section
uncertainties, the hadro-production modeling, the beam--flux variations and their convolutions with the
detectors' acceptance have been checked: the systematic error on FNR is $1-2\%$.
Evaluation of p-values for $H_0$ allows to set the possible presence of a sterile component.
However, in order to estimate the power of an experiment exclusion plots
should be also evaluated. The distortion of FNR due to a sterile neutrino component with respect to the null hypothesis
has been looked through, taking care of the correlations due to the systematic errors.
In the following both procedures are depicted.

The experiment sensitivity to the $\nu_\mu$ disappearance was evaluated by considering several estimators, 
 related either to {\em i)}  the muon produced in $\nu_\mu^{CC}$ 
or to {\em ii)} the reconstructed neutrino energy.
The muon momentum can be very effective when $H_0$ hypothesis is checked to establish the probability
of the non-sterile component, i.e. the observation of something else. In such a case the simulation is limited to the standard
processes and the FNR approach in the NESSiE environment is fully efficient.
Instead, to evaluate exclusion plots one needs to extract the oscillation parameters via Monte Carlo by looking at 
the FNR distortion in specific regions of the phase space. A new procedure  based on the
reconstructed muon momentum was also implemented to exclude regions defined by new ``effective'' variables.
Using reconstructed measured quantities allows to keep systematic errors under control.

In the second case {\em (ii)} the neutrino energy was reconstructed from  
\begin{equation}
E_\nu = \frac{E_\mu - m^2_{\mu}/(2M)}{1-(E_\mu -p_\mu\cos\theta)/M},
\end{equation}
valid in the Charge Current Quasi Elastic (CCQE) approximation,  $M$ being the nucleon mass, $E_{\mu}$ and $p_{\mu}$ the muon energy and momentum, respectively.  

We developed complex analyses to determine the sensitivity region that can be explored with an exposure of 
$6.6\times 10^{20}$ p.o.t., corresponding to 3 years of data collection on the FNAL--Booster beam.
Our guidelines were the maximal extension at small values of the mixing angle parameter and the control of the
 systematic effects.

The sensitivity of the experiment was evaluated performing three analyses that implement different techniques and approximations:
\begin{itemize}
\item method I: a Feldman\&Cousins technique (see Section V of~\cite{Fel-Cou}) with {\em ad hoc} systematic errors added to the muon momentum distribution;
\item method II: a Pearson's $\chi^2$ test~\cite{pdg} with a full correlation matrix based on full Monte Carlo simulation and reconstruction;
\item method III: a new approach based on the profile likelihoods, often referred to as modified frequentist method or CL$_s$~\cite{read-2002},
similar to that used in the Higgs boson discovery~\cite{atlas-cms}. 
\end{itemize}

Throughout the analyses the detector configuration defined in Table~\ref{tab:NFD} was considered. 


\begin{table}
\centering
\caption{\label{tab:NFD}Fiducial mass and baselines in configuration 4 for near and far detectors, used for the sensitivity analyses.}
\begin{tabular}{lcccccc}
\hline
 & Fiducial Mass (ton) & Baseline (m)\\
\hline
Near & 297 & 110\\
Far & 693 & 710\\
\hline
\end{tabular}
\end{table}



\begin{figure}[htbp]
\begin{center}
\includegraphics[scale=0.42]{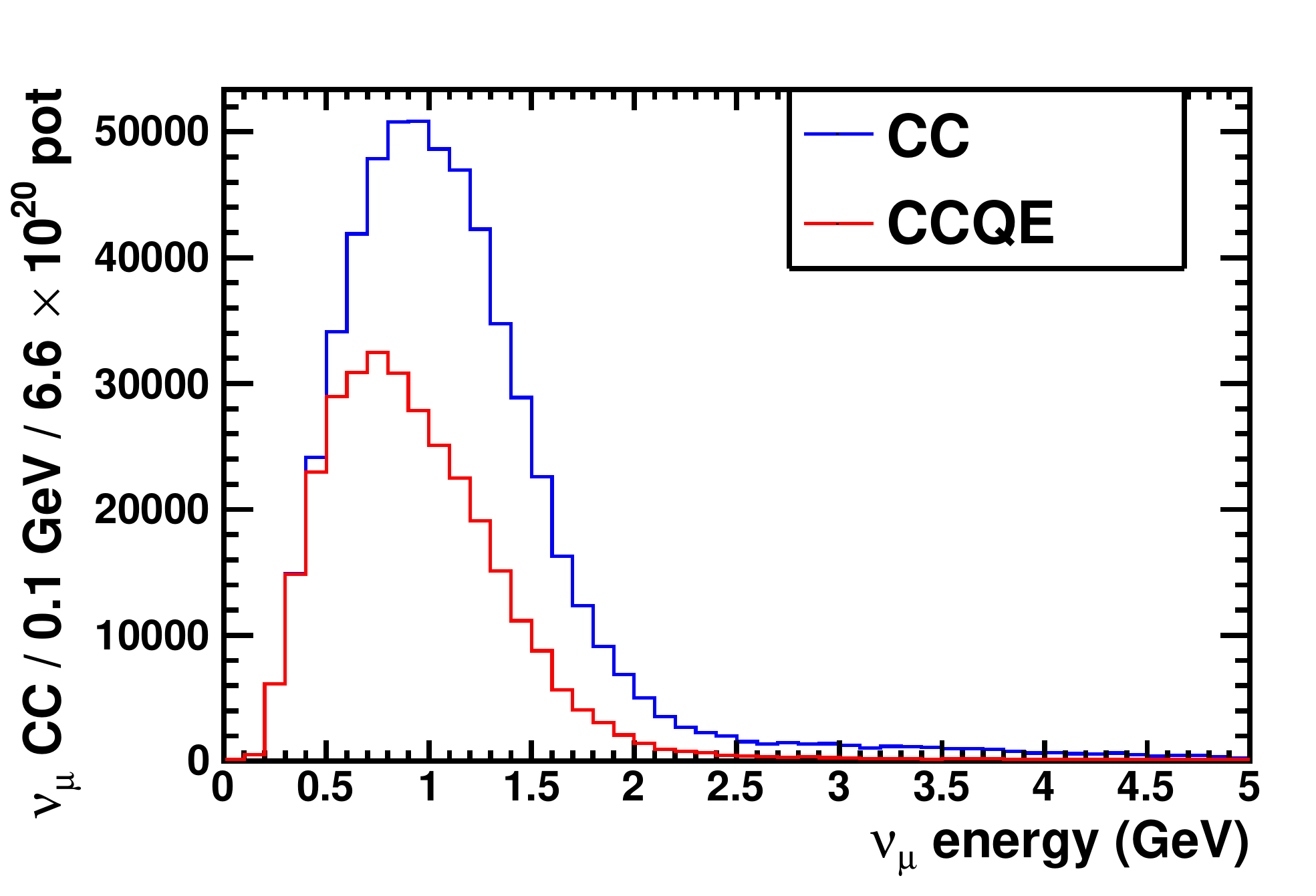}
\includegraphics[scale=0.42]{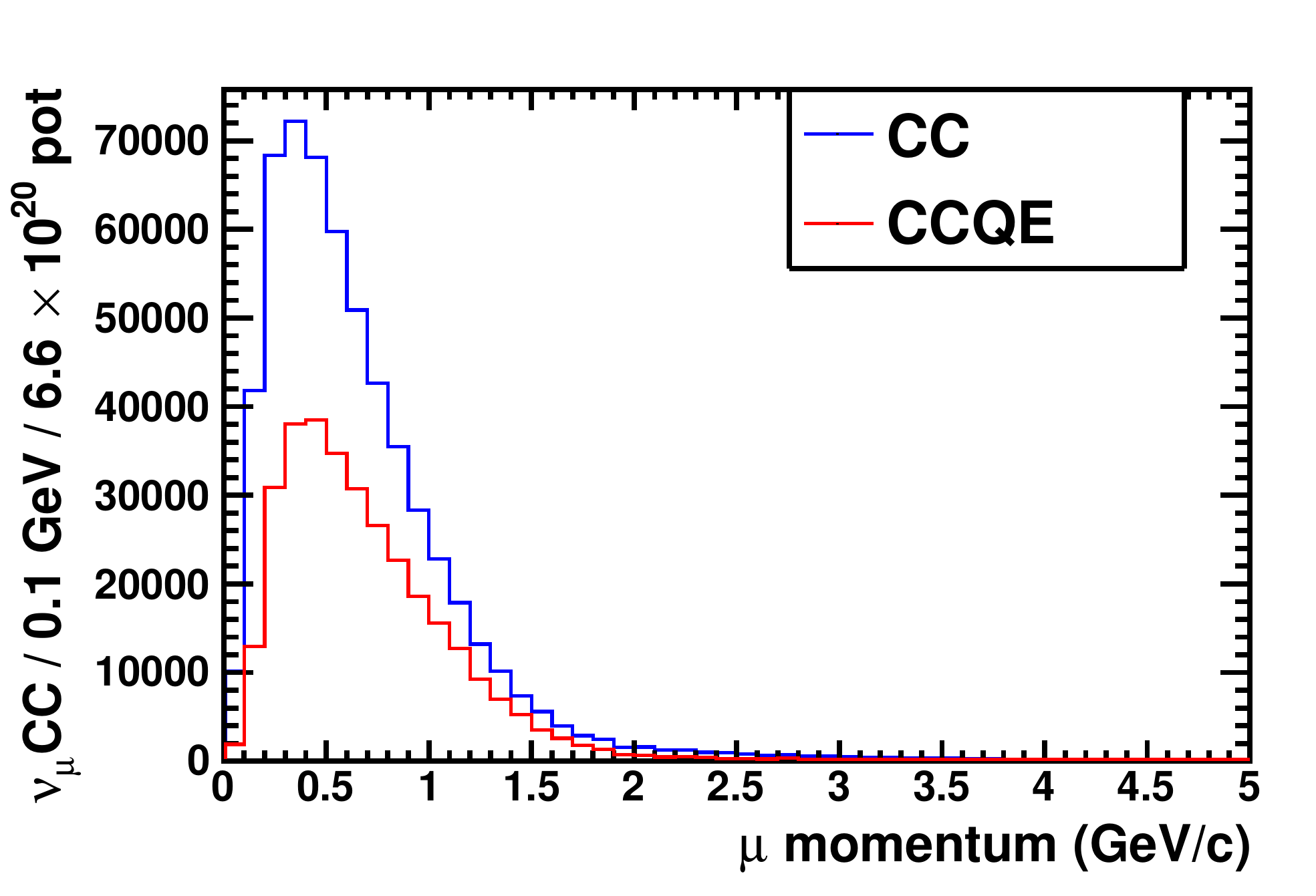}
\caption{\label{fig:norma-interac}(color online) The total number of expected \numu CC interactions seen by the Far detector at 710 m, as a function of 
$E_{\nu}$ (top) and  $p_{\mu}$ (bottom), for the expected
luminosity in 3 years of data taking ($6.6\times 10^{20}$ p.o.t.) with the FNAL--Booster beam in positive--mode running.
The sub--sample corresponding to the CCQE component is also shown.}
\end{center}
\end{figure}

The distributions of events, either in  $E_{\nu}$ or  $p_{\mu}$, normalized to the expected
luminosity in 3 years of data taking ($6.6\times 10^{20}$ p.o.t.) with the FNAL--Booster beam running in positive focusing mode, are reported in Fig.~\ref{fig:norma-interac}. 

The study of the \nubarmu disappearance is reported in 
Section~\ref{sec:antinu} with results obtained from method III.


\subsection{Sensitivity Analyses}

In the three analyses the two--flavour neutrino mixing in the approximation of one mass dominance was considered.
The oscillation probability is  given by:

\begin{equation}
\label{eq:2flavour}
P({\nu_\mu\to\nu_\mu}) = 1 - \sin^{2}(2\theta_{new})\sin^{2}\left (\frac{1.27\Delta m^{2}_{new}L[{\rm km}]}{E[{\rm GeV}]}\right ),
\end{equation}
where $\Delta m^{2}_{new}$ is the mass splitting between a new heavy--neutrino mass--state and the
heaviest among the three SM neutrinos, and $\theta_{new}$ is the corresponding effective mixing angle.

In the selected procedures (Feldman\&Cousins approach, $\chi^2$ test with Near--Far correlation matrix and CL$_s$ profile likelihoods)  the evaluation of the sensitivity region to sterile neutrinos
was computed at  95\% C.L. Some recent use of more stringent Confidence Limits (even to 10 $\sigma$'s~\cite{nustorm}) 
was judged un--necessary, provided the correct and conservative estimation of the systematic errors. Besides, we note that the measurement of muon tracks is a quite old
and proven technique with respect to the more difficult detection and measurement of electron--neutrino interactions in Liquid--Argon systems.


\subsubsection{Method I (Feldman\&Cousins technique)}

In  method I the far--to--near ratio was written as $R_i = F_i/(kN_i)$, where $F_i$ and $N_i$ are the number of events in the $i$--th bin of the muon--momentum distribution in the far and near detectors, respectively, and $k$ is a bin--independent constant factor used to normalize each other the near and far distributions. 
For each value of the oscillation parameters, $\sin^22\theta_{new}$ and $\Delta m^2_{new}$, the $\chi^2$ is computed as
\begin{equation}
\chi^2 = \sum_{i=1}^N\left (\frac{1-R_i/R_{0,i}}{\sigma_{R_{0,i}}} \right )^2,
\end{equation}
where $R_{0,i}$ is the far--to--near ratio in absence of oscillation and $\sigma_{R_{0,i}}$ is the quadratic sum of the statistical error and a fixed, bin--to--bin uncorrelated, systematic error. In the Feldman\&Cousins approach 
a $\Delta\chi^2_{cut}(\sin^22\theta_{new},$ $\Delta m^2_{new})=\chi^2(\sin^22\theta_{new}, \Delta m^2_{new})-\chi^2_{min}$ cut is applied.
For every ($\sin^22\theta_{new}$, $\Delta m^2_{new}$) oscillated spectra were generated and
fitted to obtain  the $\chi^2_{min}$. The distribution of $\Delta
\chi^2(\sin^22\theta_{new}, \Delta m^2_{new})$ is cut at 95\% to define the  $(\sin^22\theta_{new}, \Delta m^2_{new})$ exclusion region. 
The critical value on $\Delta\chi^2_{cut}$ can be determined by either sampling the $\Delta\chi^2$ distribution as in Feldman\&Cousins
or by applying the standard $\chi^2_{cut}$ fixed--value for the 95\% C.L. and two degrees--of--freedom.
It was verified that in both cases the obtained results are very similar in the whole ($\sin^22\theta_{new}$, $\Delta m^2_{new}$) space.

\begin{figure}[htbp]
\centering
\includegraphics[scale=0.44,type=pdf,ext=.pdf,read=.pdf]{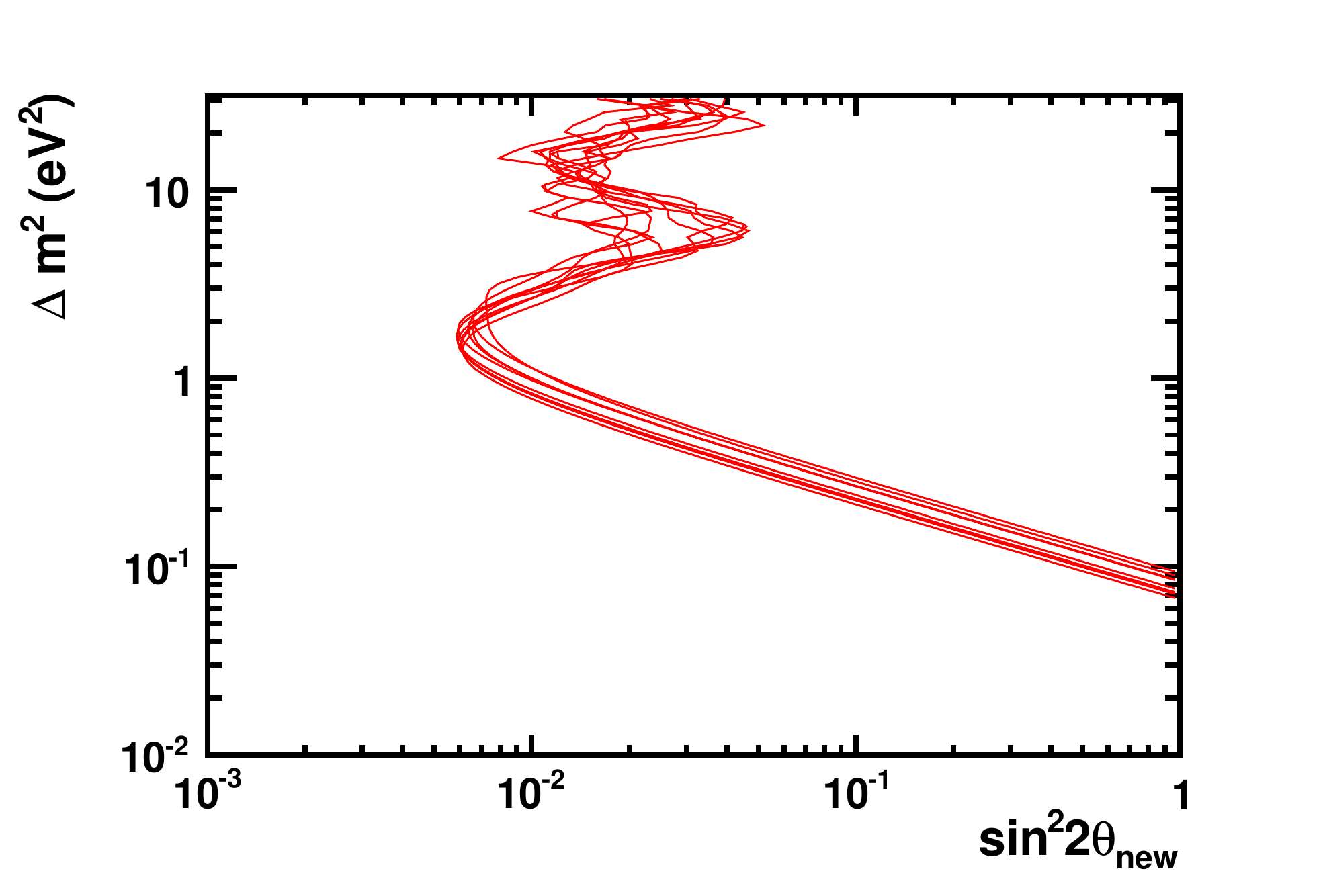}
\includegraphics[scale=0.44,type=pdf,ext=.pdf,read=.pdf]{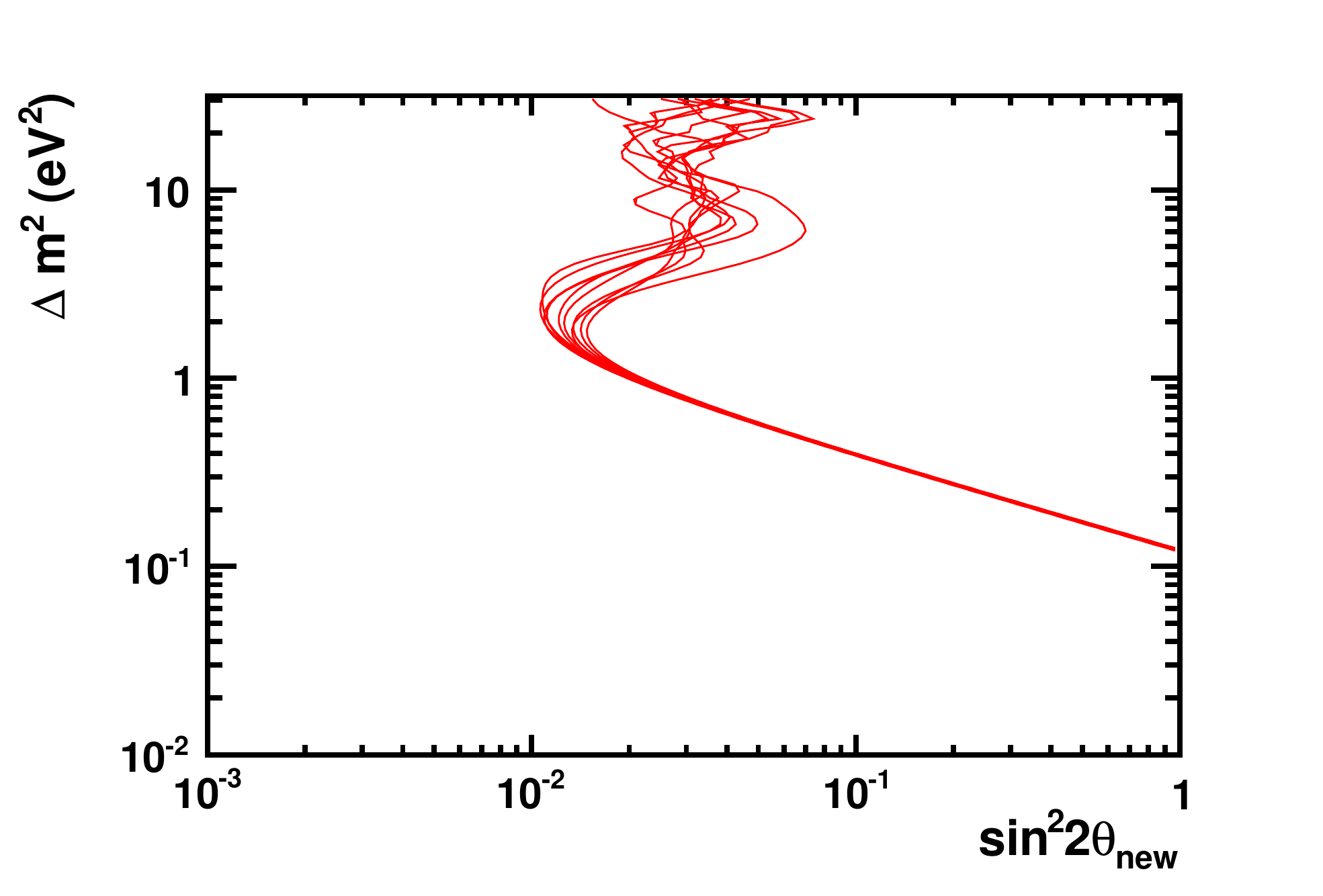}
\caption{\label{fig:sens0}Top: sensitivity curves at 95\% C.L.  with ten simulated toy experiments 
and no systematic uncertainties.  In the fit procedure $p_\mu$ was used as observable with a cut of 500~MeV/c. 
10$^4$ sampling points, uniformly distributed in log scale, were generated. 
The expected integrated luminosity for 3 years of data taking ($6.6\times 10^{20}$ p.o.t.) with the FNAL--Booster beam in positive--mode 
running 
was assumed.
Bottom: as above but using a bin--to--bin uncorrelated systematic--error $\epsilon_{sys} = 0.01$.}
\end{figure}

Results are shown in Fig.~\ref{fig:sens0} for a set of ten simulated null experiments.
In the top plot a systematic error $\epsilon_{sys}=0$  was used.
In the bottom plot a bin--to--bin uncorrelated systematic--error
 $\epsilon_{sys}=0.01$ was assumed (see Section 12.1 of~\cite{nessie-fnal} for more details).


\subsubsection{Method II ($\chi^2$ test with Near--Far correlation matrix)}
In method II the sensitivity to the $\nu_\mu$ disappearance was evaluated using 
two different observables, the muon {\em range} and the {\em number of crossed RPC planes}. 
The correlations between the data collected in the far and near detectors are taken into account through the covariant matrix of the observables. 
The $\chi^2$ is given by 
\begin{equation}
\chi^2 = \sum_{i=1}^{N}\sum_{j=0}^N\left (kN_{i} - F_i\right ) \left (M^{-1}\right )_{ij} \left (kN_{j} - F_j\right ),
\end{equation}
where $M$ is the covariance matrix~\cite{covM} of the uncertainties (statistical and  bin--to--bin systematic correlations~\cite{covM2}). 

The $\nu_{\mu}$ disappearance can be observed either by a deficit of events ({\em normalization}) or, also, by a distortion of the observable spectrum 
({\em shape}\footnote{Note that the shape analysis looks at the same distributions of Method I.}), 
which are affected by systematic uncertainties expressed by the normalization error--matrix and the shape error--matrix,
respectively. The {\em shape} error--matrix represents a migration of events across the bins. In this case the uncertainties are associated with changes
not affecting the total number of events. Consequently, a depletion of events in some region of the spectrum should be compensated by an enhancement in others. 
Details of the model used for the shape error--matrix can be found in~\cite{nessie-fnal}.   

The distributions of the muon {\em range} and of the {\em number of crossed planes} were computed using GLoBES~\cite{GLoBES} with the smearing matrices obtained by the full Monte Carlo simulation described in Section~\ref{sec:spect1}.

%

By applying the frequentist method the $\chi^{2}$ statistic distribution was looked at  
in order to compute the sensitivity to oscillation parameters. 
Different cuts on the range and on the number of crossed planes were studied. Furthermore sensitivity plots were computed by introducing bin--to--bin correlated systematic uncertainties 
by considering either 1\% correlated error in the normalization or alternatively 1\% correlated error in the spectrum shape. 

\begin{figure}[htbp]
\centering
{\includegraphics[scale=0.43]{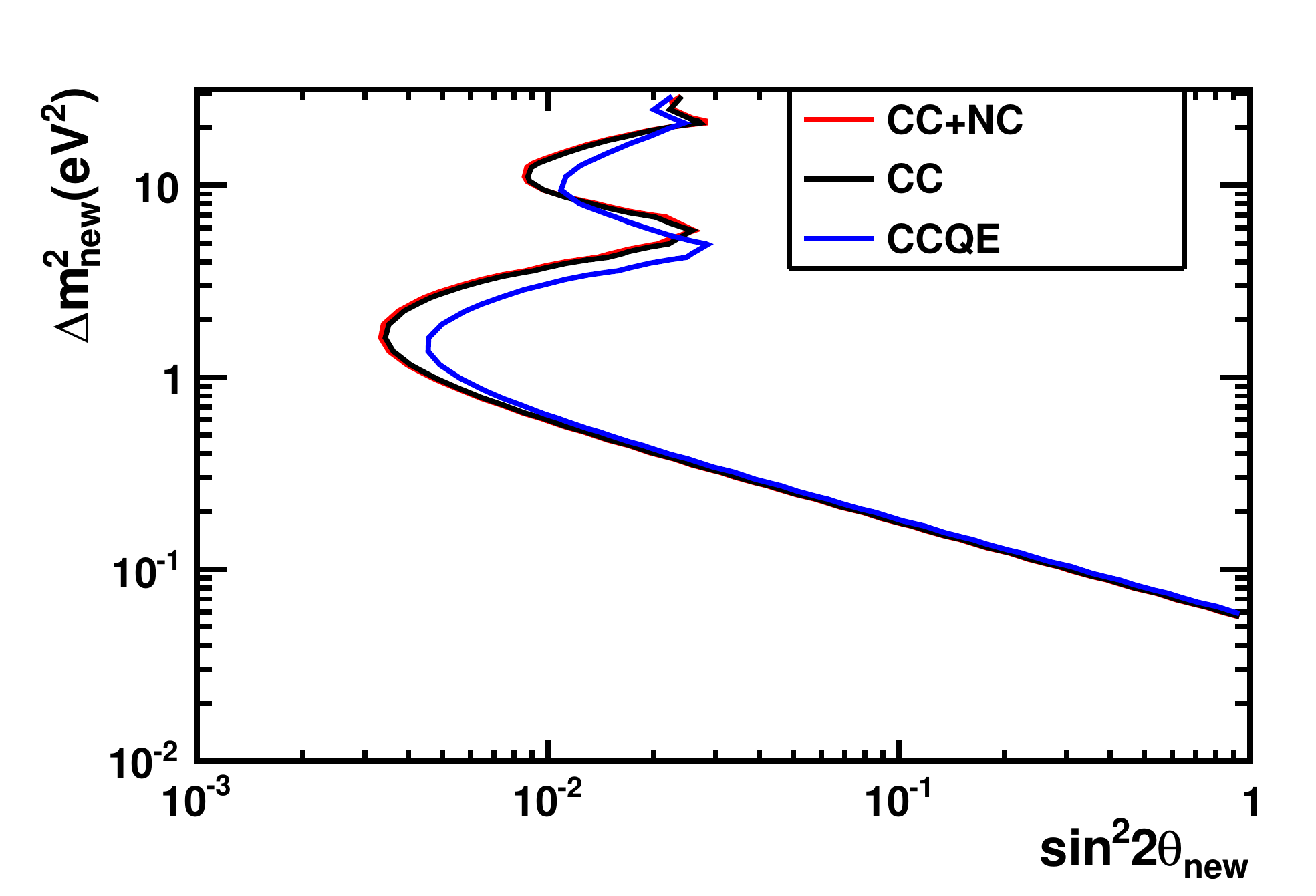}} 
\caption{\label{fig:sensitivity_plot_all_corrShape}(color online) 95\% C.L. sensitivity obtained using the muon {\em range} for all the
interaction processes  (Quasi Elastic, Resonant, Deep Inelastic Scatering): CC (black), CC+NC (red) and for  
CCQE events only (blue). 1\% bin--to--bin correlated error in the {\em shape} is considered.
The expected integrated luminosity for 3 years of data taking ($6.6\times 10^{20}$ p.o.t.) with the FNAL--Booster beam in positive--mode
running 
was assumed.}
\end{figure}

As a representative result the sensitivity computed using the {\em range} as observable and taking the 1\% correlated error 
in the shape, is plotted in Fig.~\ref{fig:sensitivity_plot_all_corrShape}. 
Instead the normalization correlated--error would slightly reduce the
 sensitivity region around $\Delta m^2_{new}=1$ eV$^2$.
Moreover, the sensitivity region obtained by using the sum of CC and NC events is  almost the same
 as that  obtained with CC events only (see Section 12.2 in~\cite{nessie-fnal}). That proves that the result is not
 affected by the NC background events.


\subsubsection{Method III (CL$_s$ profile likelihoods)}\label{sec:cls}
In the profile CL$_s$ method we introduce a new test--statistics that depends on a {\em signal--strength} variable. 
By looking at Eq.~\ref{eq:2flavour}  the factor $\sin^22\theta_{new}$ acts as an amplification 
quantity for a fixed $\Delta m^{2}_{new}$.
Therefore a signal--strength $\mu$ can be identified  with $\sin^{2}2\theta_{new}$
to construct the estimator function:
\begin{equation}\label{eq:signal-strength}
f = \frac{1-\mu\cdot\sin^{2}(1.27\ \Delta m^{2}_{new}\ L_{Far}/E_{\nu})}{1-\mu\cdot\sin^{2}(1.27\ \Delta m^{2}_{new}\ L_{Near}/E_{\nu})}.
\end{equation}

In a simplified way, for each $\Delta m^{2}_{new}$, a sensitivity limit on $\mu$ can be obtained from the {\em p--value} of the distribution of the estimator $f$
in Eq.~\ref{eq:signal-strength}, in the assumption of background--only hypothesis. 

That procedure does not correspond to
computing the exclusion region of a signal, even if it provides confidence for it. The exclusion plot should be obtained by fully
developing the CL$_s$ procedure as described in Section 12.3 of~\cite{nessie-fnal}. However, since we are here mainly interested in
exploiting the sensitivity of the experiment, the procedure provides already insights into that. 
Its result comes fully compatible with the previous two analyses, which follow the usual neutrino analyses found in the literature.

Moreover, following the same attitude, an even more {\em aggressive} procedure can be applied. Since the deconvolution
from $p_{\mu}$ to $E_{\nu}$ introduces a reduction of the information, we investigated whether the more
direct and measurable parameter, $p_{\mu}$, can be a valuable one. In such a case Eq.~\ref{eq:signal-strength} becomes:
\begin{equation}\label{eq:signal-strength-mu}
f = \frac{1-\mu\cdot\sin^{2}(1.27\ \Delta m^{2}_{new}\ L_{Far}/p_{\mu})}{1-\mu\cdot\sin^{2}(1.27\ \Delta m^{2}_{new}\ L_{Near}/p_{\mu})}
\end{equation}

\begin{figure}[htbp]
\centering
\includegraphics[scale=0.43]{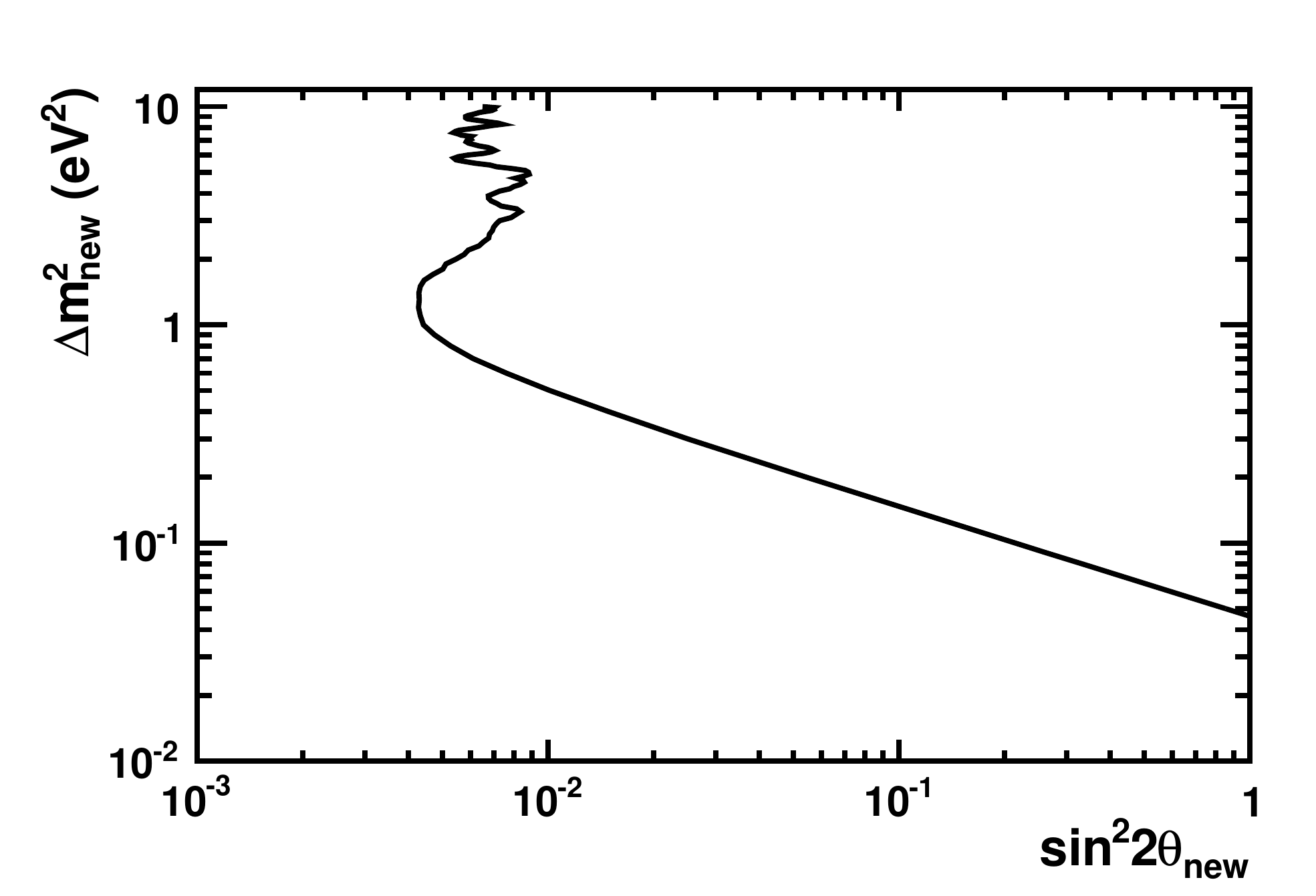}
\caption{\label{fig:exclu-pmu}The sensitivity at 95\% of C.L. obtained by computing the modified raster--scan method, 
in a CL$_s$ framework, and by
using the reconstructed muon momentum as estimator (equation~(\ref{eq:signal-strength-mu})). 
The expected integrated luminosity for 3 years of data taking ($6.6\times 10^{20}$ p.o.t.) with the FNAL--Booster beam in positive--mode
running 
was assumed.
A conservative cut of $p_{\mu,rec}\ge\ 500$~MeV/c was applied.}
\end{figure}

The corresponding sensitivity plot is shown in Fig.~\ref{fig:exclu-pmu}. It provides an ``effective'' sensitivity limit in the ``effective'' variables
$\Delta m^{2}$  and the reconstructed muon momentum, $p_{\mu,rec}$. 
By applying the Monte Carlo deconvolution from $p_{\mu,rec}$ to $E_{\nu}$
we checked that the ``effective'' $\Delta m^{2}$ is simply scaled--off towards lower values, not affecting the mixing angle limit\footnote{ 
The scaled-off feature is evident from the comparisons of the sensitivity curves in the two cases, either $E_{\nu}$ or $p_{\mu,rec}$ (Figs.~51 and 52 of the original proposal~\cite{nessie-fnal}).}. 
The merits of the $p_{\mu,rec}$ are multiple: it is not affected as $E_{\nu}$ by the propagation error due to 
the deconvolution process, the systematics errors due to the reconstruction of the events (efficiency, acceptance, background)
are directly included since it corresponds to a measured quantity, the estimator is  powerful  in  identifying
a possible new signal/anomaly.


\subsection{Conclusions for Section~\ref{sec:analysis}}

\begin{figure}[htbp]
\centering
\includegraphics[scale=0.43]{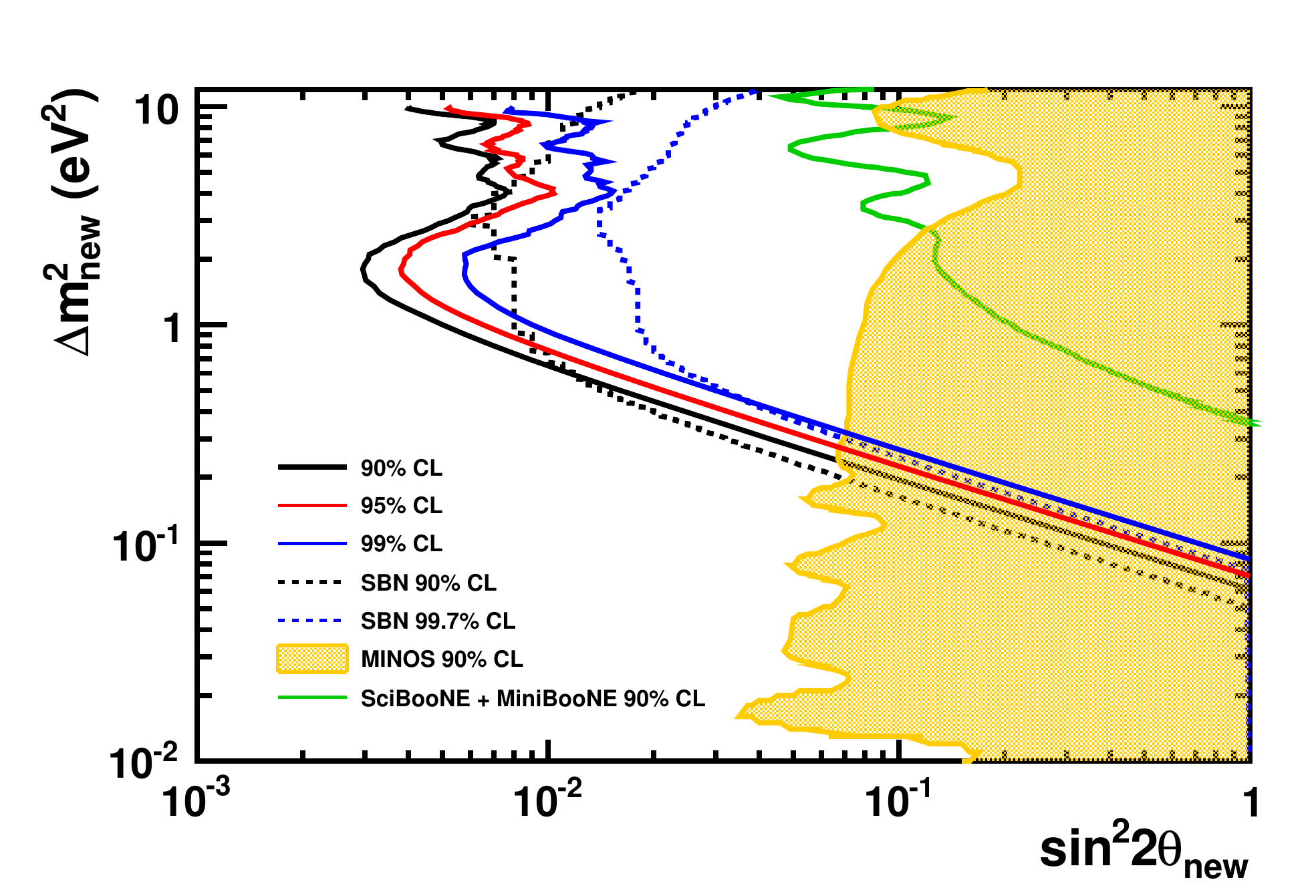}
\caption{\label{fig:exclu-numu-final}The sensitivity curves obtained using the modified raster--scan method (equation~(\ref{eq:signal-strength})), 
in a CL$_s$ framework, for 90\%, 95\% and 99.7\% C.L., for an
expected integrated luminosity of 3 years of data taking ($6.6\times 10^{20}$ p.o.t.) with the FNAL--Booster beam in positive--mode
running.
An uncorrelated 1\% systematic error and a  conservative cut of $p_{\mu,rec}\ge\ 500$~MeV/c were used. 
The filled area corresponds to the MINOS~\cite{recent-MINOS} result and the green curve to the
 MiniBooNE/SciBooNE~\cite{mini-sci-mu} limit (at 90\% C.L.).
The two dashed lines corresponds to the sensitivity predicted by the new SBN proposal ~\cite{SBN}, at 90\% and 99.7\% C.L.}
\end{figure}

The sensitivity curves obtained with different analyses prove the possibility to explore a very large region in the
mass--scale and mixing--angle plane, larger than other current proposals. Using a configuration with two (massive) detectors, 
one at 110 m on--axis, and one at 710 m off-axis (configuration 4, see Table~\ref{tab:NFD}), the achievable sensitivity curves are drawn
in Fig.~\ref{fig:exclu-numu-final} for several C.L., compared to existing limits~\cite{mini-sci-mu,recent-MINOS} 
and the predicted sensitivities of the SBN project~\cite{SBN}. 
A systematic error of 1\% has been assumed and a conservative cut
$p_{\mu,rec}\ge\ 500$~MeV/c was applied.  A sensitivity to mixing angles 
below $10^{-2}$ in $\sin^22\theta_{new}$ 
can be obtained in a large region of $\Delta m^{2}$ around $1\ {\rm eV}^2$ scale. 

It is  noted that by applying more elaborated reconstruction algorithms than those used in the present analysis the 
500 MeV/c cut in $p_{\mu,rec}$ could be lowered to $200-300$ MeV/c. 
The exclusion region could then be significantly extended 
to $\Delta m^{2}<1$ eV$^2$, even if systematic errors should generally be larger and therefore detailed studies would be required
to really access that limit.

We demonstrated that a sophisticated statistical tool (method III) can be applied to get hints of new neutrino states at lower
mass--scale than that achievable with the usual ones (methods I and II), by making use of a different estimator (the reconstructed muon momentum), less dependent of the Monte Carlo simulation. Thus we conclude that, on top of the exclusion limits, 
a robust confidence is accomplished on the identification of a possible new signal.

\section{Sensitivity for the antineutrino disappearance}\label{sec:antinu}
\begin{figure}[htbp]
\begin{center}
\includegraphics[scale=0.43]{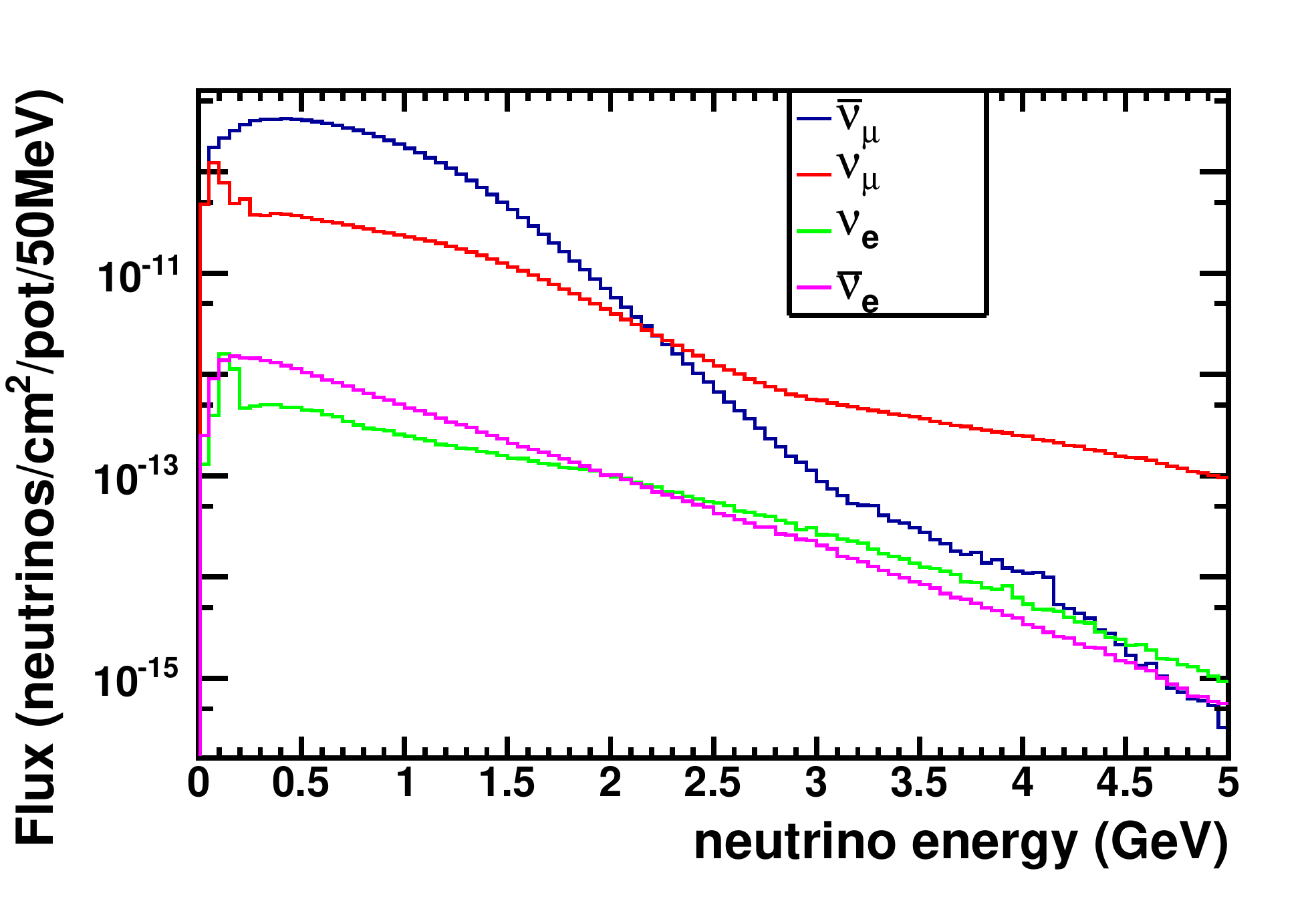}
\caption{\label{fig:BNB_NegPol}(color online) The flux components of the Booster antineutrino beam (from~\cite{G4BNBflux}).}
\end{center}
\end{figure}

In negative--focusing mode the Booster beam contains a large neutrino component. In terms of flux the contamination amounts to
 15\% at 1 GeV, 30\% at 1.5 GeV and surpasses the antineutrino flux above 2 GeV (see Fig.~\ref{fig:BNB_NegPol}). 
In this energy region the measurement of the charge 
on event--by--event basis is an efficient tool. 
Although a comprehensive study of the spectrometers' ultimate performance goes beyond the scope of this paper,  we   estimated the improvement on the final sensitivity by using their charge ID capability. We applied Method III of Section~\ref{sec:cls}, under some additional assumptions on the neutrino--anti\-neu\-trino components.
The contamination of the neutrino events resulting from the charge mis--ID probability ($\eta$)
was included in 
an uncorrelated way to the near and far data samples, with/without the assumption that only antineutrinos oscillate.
The assumption that the oscillation phenomenon acts differently for neutrinos and antineutrinos 
puts limits on their possible correlated oscillation probabilities. Though the resulting sensitivity curves correspond to the worst case scenario, they  
provide a decoupled insight into the possible
measurement of the antineutrino disappearance at 1 eV mass--scale and small mixing angle, for the first time.

For the antineutrino disappearance search two main results exist, an old one by the CCFR experiment~\cite{ccfr-anti}
and more recent ones from the MiniBooNE~\cite{mini-mu} and MiniBooNE/SciBooNE~\cite{mini-sci-antimu}. 
The MiniBooNE results come with a 20\%--25\% contamination 
of the intrinsic neutrino flux and were obtained assuming
that only antineutrinos oscillate while neutrinos do not.

\begin{figure}[htbp]
\begin{center}
\includegraphics[scale=0.42]{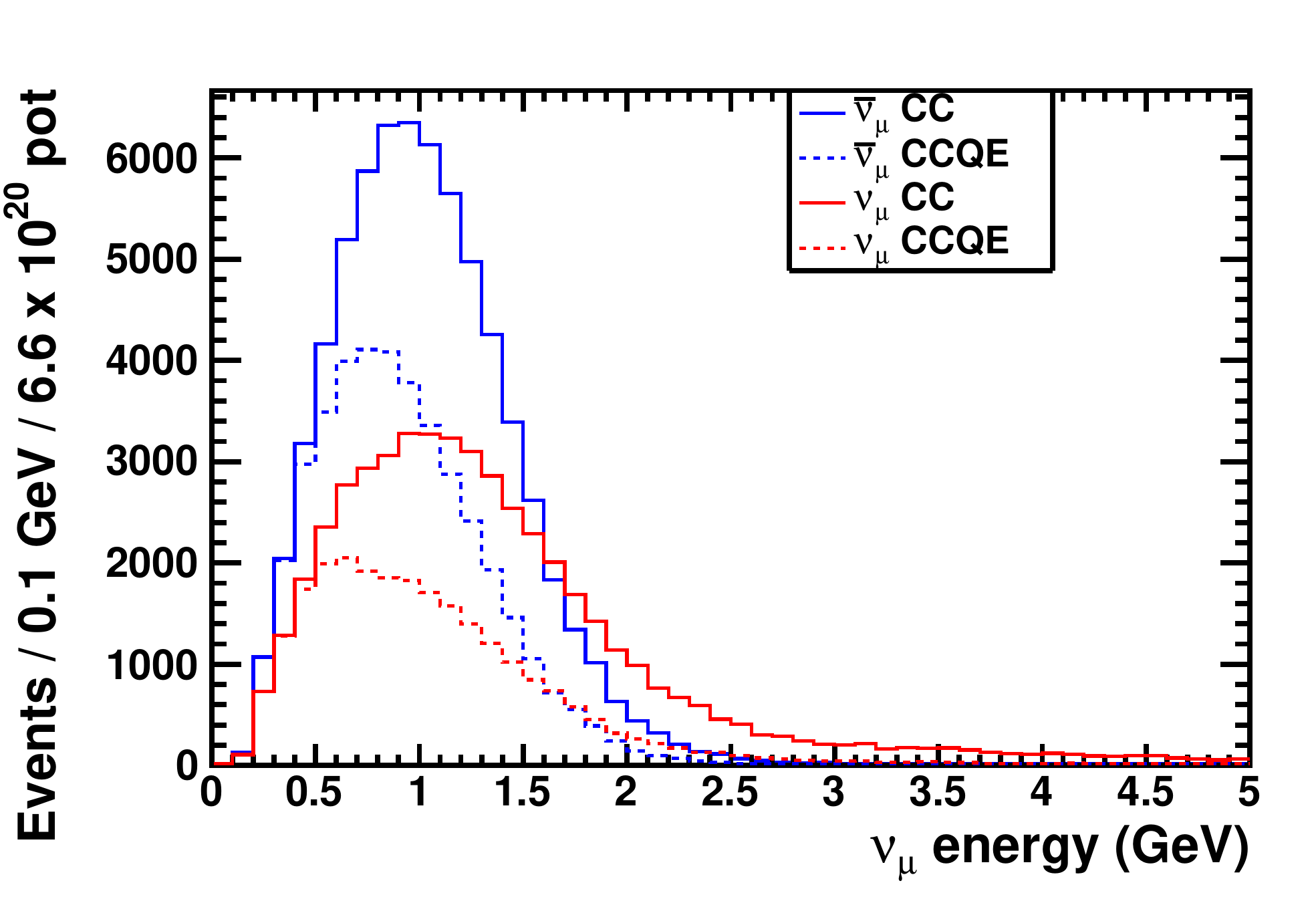}
\includegraphics[scale=0.42]{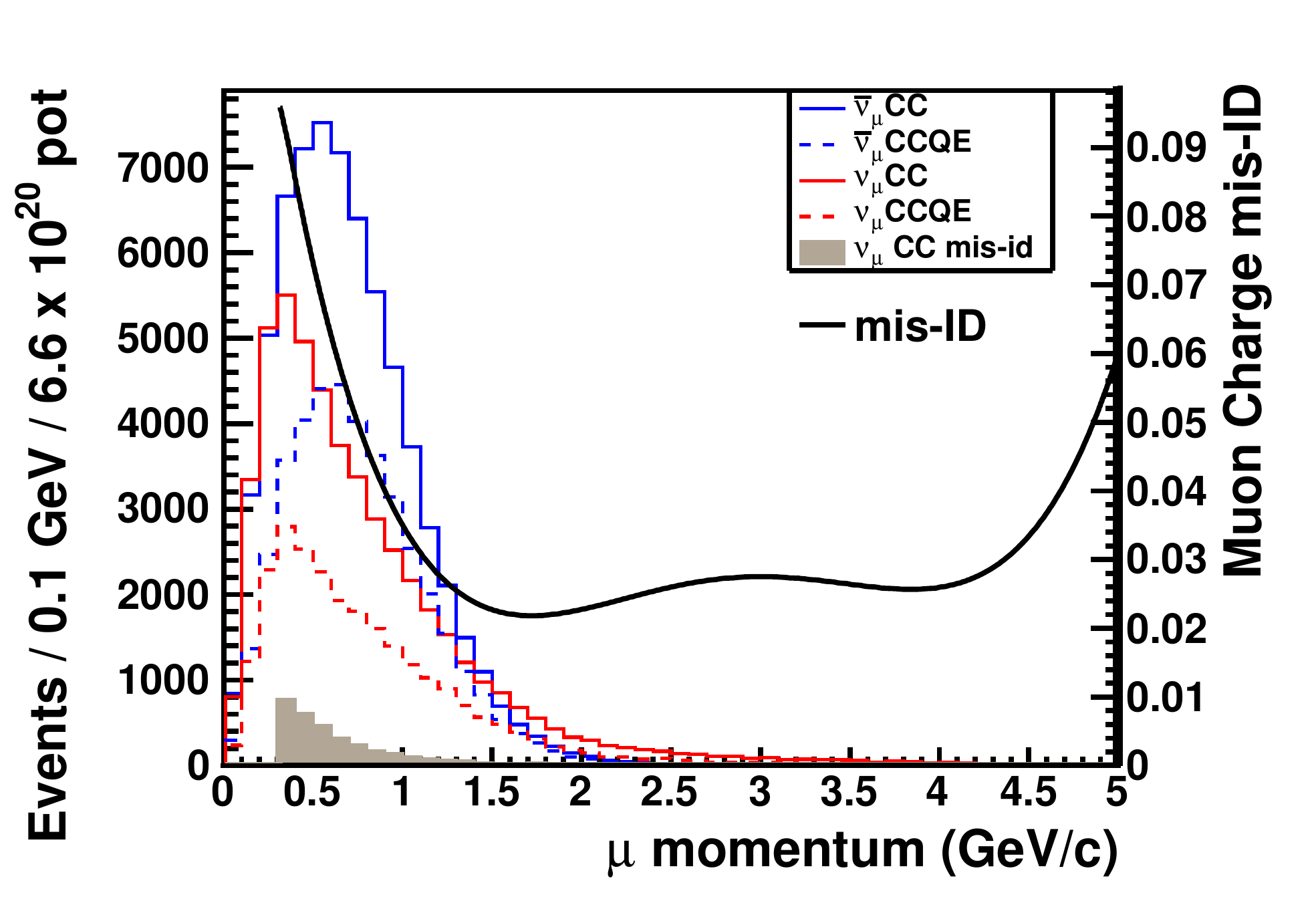}
\caption{\label{fig:norma-interac-anti}(color online) The total number of foreseen \nubarmu CC interactions seen by the Far detector at 710 m, as function of the
$E_{\nu}$ (top) and the $p_{\mu}$ (bottom), for the expected
luminosity in 3 years of data taking ($6.6\times 10^{20}$ p.o.t.) with the FNAL--Booster beam in negative--mode
running. Separate antineutrino and neutrino data are shown, with also the sub--samples corresponding to the CCQE component (dashed histograms).
In the bottom plot the mis--ID of the muon charge as provided by the spectrometers, is overlaid (black curve and left scale)~\cite{bopera}. The grey zone in the same plot corresponds to the expected CC neutrino contamination once measured the muon
track (integrated 3200 over 58600 CC events in the 0.3 -- 5 GeV interval).}
\end{center}
\end{figure}

A total integrated luminosity corresponding to 3 years of running at the Booster in negative--focusing mode was  considered.
The number of events that could be collected at the far detector is displayed
in Fig.~\ref{fig:norma-interac-anti}. The neutrino sub--component is highly enhanced because of its larger cross-section
with respect to the antineutrino one.

\begin{figure}[htbp]
\begin{center}
\includegraphics[scale=0.43]{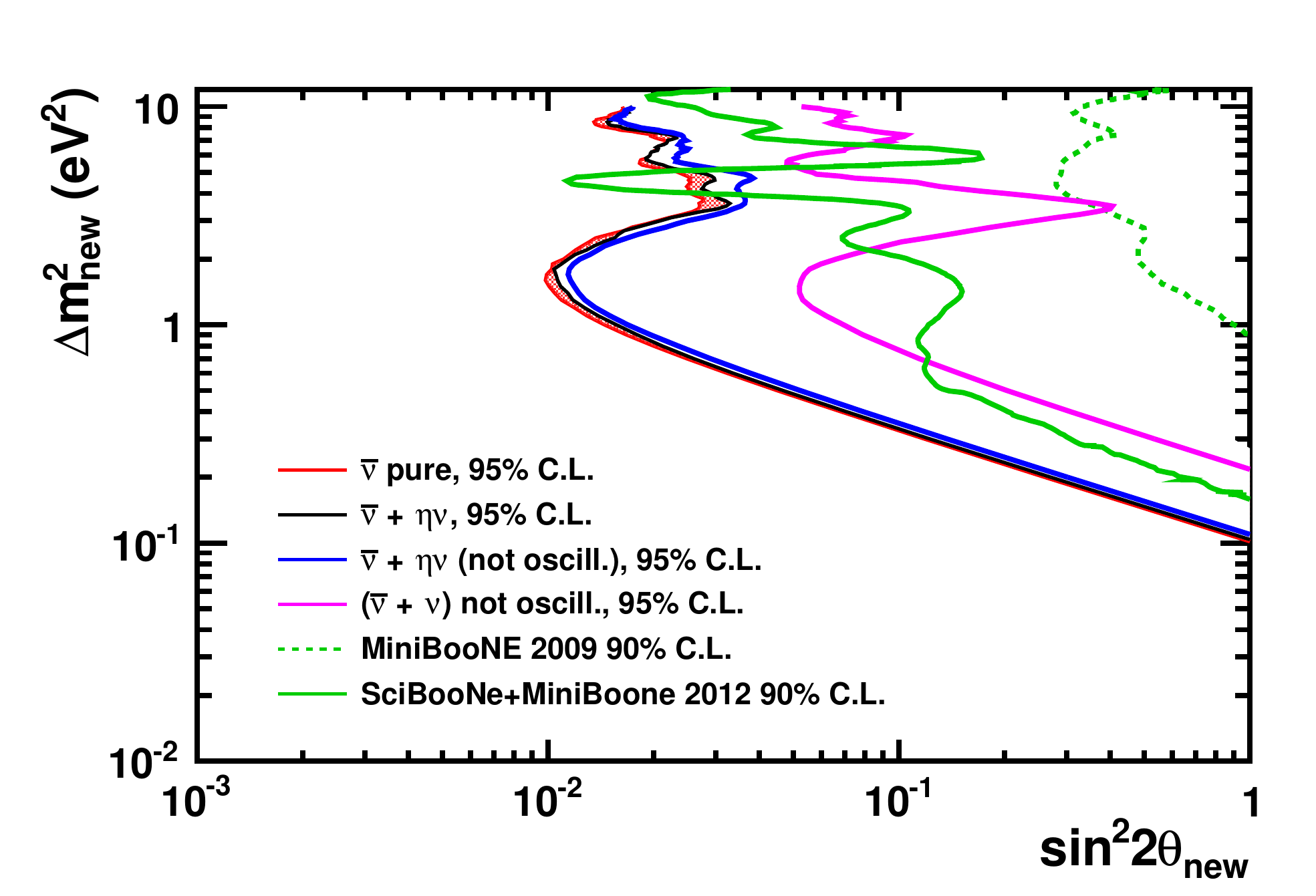}
\caption{\label{fig:limits-antinu}(color online) The four sensitivity curves (see text for explanation) obtained using the modified raster--scan method (equation~(\ref{eq:signal-strength})), 
in a CL$_s$ framework, for 95\% C.L., for an
expected integrated luminosity of 3 years of data taking ($6.6\times 10^{20}$ p.o.t.) with the FNAL--Booster beam 
 running in negative--focusing mode. $\eta$ stays for the charge mis--ID probability.
An uncorrelated 1\% systematic error and a  conservative cut of $p_{\mu,rec}\ge\ 500$~MeV/c were used. 
Previous results are also shown.
The {\em dashed green curve} corresponds to the 90\% C.L MiniBooNE~\cite{mini-mu} limit while the 
{\em plain green curve} corresponds to its improved analysis and data collection with also the SciBooNE data~\cite{mini-sci-antimu} .}
\end{center}
\end{figure}

To evaluate our sensitivity to the {\em signal--strength} estimator (Section~\ref{sec:cls})
a set of four samples has been considered, under different assumptions:
\begin{description}
\item[1)] the pure anti-neutrino sample (perfect rejection of neutrino contamination);
\item[2)] the anti-neutrino sample with the neutrino contamination as determined by the spectrometers charge mis-ID. The same oscillation law is assumed for neutrinos and anti-neutrinos;
\item[3)] as above, but assuming no oscillation for the neutrino contamination;
\item[4)] the anti-neutrinos sample with full neutrino contamination (no charge ID), assuming no oscillation for the neutrino contamination.
\end{description}
Fig.~\ref{fig:limits-antinu} shows the corresponding sensitivity curves, that have less and less power going from
assumption (1) to (4), as expected.
The most sensitive curve (case 1 and {\em red line}) corresponds to the pure sample of antineutrinos. The {\em black line} for case (2) is obtained 
by including  the muon charge mis-ID in the collected event sample and increasing correspondingly the statistical errors associated to the neutrino and anti-neutrino components.
The curve for case 3 ({\em blue line}) is obtained by assuming that the neutrino component, identified by the muon charge, do not oscillate, 
then decreasing the total sample while contributing to the statistical error. The {\em purple line} for (4) indicates the sensitivity in case
the neutrino component is not identified and assumed not oscillating. 

The key feature of the charge identification is apparent since the quite small contamination coming from the mis--ID
produces small corrections. In contrast lacking of charge measurement oblige to assume an oscillation
pattern for the contamination component and reduces drastically the amount of the equivalent statistical sample
and the sensitivity region (case 4).

\section{Conclusions}\label{conclu}
Existing {\em anomalies} in the neutrino sector  may hint to the existence of  one or more additional {\em sterile} neutrino states. 
A detailed study of the physics case was performed to set up a Short--Baseline experiment at the FNAL--Booster neutrino
beam exploiting the study of the muon--neutrino charged--current interactions. An independent measurement on \numu, complementary to
the already proposed experiments on \nue, is mandatory
to either prove or reject the existence of sterile neutrinos, even in case of null result for \nue.
Moreover, very massive detectors are mandatory to collect a large number of events and therefore improve the disentangling of systematic effects.

The best option in terms of physics reach and funding constraints is provided by two spectrometers based on dipoles iron magnets,
at the Near and Far sites, located at 110 (on--axis) and 710~m (on surface, off--axis) from the FNAL--Booster neutrino source, respectively, 
possibly placed behind the proposed LAr detectors. 

A full re-use of the OPERA spectrometers, when dismantled, would be feasible.
Each site at FNAL can host a part of the two coupled OPERA magnets, based on well know technology,
allowing to realize ``clone'' detectors at the Near and Far sites.
The spectrometers would be equipped with RPC detectors, already 
available, which have demonstrated their robustness and effectiveness.

With that configuration one would succeed in keeping the systematic error at the level of $1-2$\% for the measurements of the 
\numu interactions, i.e. the measurement of the muon--momentum at the percent level and the identification
of its charge on event--by--event basis, extended to well below 1 GeV.

The achieved sensitivity on the mixing angle between the standard neutrinos and a new state is well below 0.01 for the
\numu mode. The measurement of the muon charge on event--by--event basis has been demonstrated to be very efficient for the estimation
of possible disappearance antineutrino phenomena, for the first time at the level of few percents for the
mixing angle and a
mass scale around 1 eV.


\section*{Acknowledgements}
\label{sec-ack}

We wish to thank the indications and encouragements received by the European Strategy Group and P5 committees, as well as by CERN and FNAL through their directorates. 
We are indebted to INFN for the continuous support all along the presented study. We would also like to thank  
the Russian Program of Support of Leading Schools (grant no. 3110.2014.2) and the Program of the Presidium of the Russian Academy of Sciences ``Neutrino physics and Experimental and theoretical researches of fundamental interactions connected with work on the accelerator of CERN''.
We acknowledge the precious collaboration of our colleagues of the technical staff, in particular L. Degli Esposti, C. Fanin, R. Giacomelli, C. Guandalini, A. Mengucci and M. Ventura.
The contributions of A. Bertolin, M. Laveder, M. Mezzetto and M. Sioli are also warmly acknowledged.
Finally, we are forever indebted to our friend and colleague G. Giacomelli, whose contribution and support we deeply miss.






\end{document}